%% file: Inf_Coalition_min_Sym_with_full_Appendix.tex
\documentclass[10pt, conference, letterpaper]{IEEEtran}

\usepackage{enumerate}
\usepackage{lineno}
\usepackage{amssymb}
\usepackage{amsmath}
\usepackage{adjustbox}
\usepackage{bbm}
\usepackage{color}
\usepackage{subfiles}

\newcommand{\eop} {\hfill{$\blacksquare$}}
\newcommand{\Cmnt}[1] {}
\newtheorem{theorem}{Theorem}
\newtheorem{corollary}{Corollary}

\newtheorem{lemma}{Lemma}

\newcommand{\AsymCmnt}[1]{}   % Just for commenting out Asymmetric player results as of now..

\newcommand{\TR}[2]{#2}  % {#1}  For TR and {#2}  for Paper 
\newcommand{\ignore}[1]{  }

\begin{document}

\title{ Cooperative Resource Sharing with Adamant Player} 
 \author{
Shiksha Singhal and  Veeraruna Kavitha, \\
IEOR, Indian Institute of Technology Bombay, India
 }

%\institute{Stockholm University, Stockholm, Sweden\\\email{elena.touli@math.su.se}\andthe Ohio State University Columbus, Ohio, U.S.A.\\\email{yusu@cse.ohio-state.edu}\\}\\

%\authorrunning{Mokhov, Sutcliffe and Voronkov}

% \title{FPT-Algorithms for computing Gromov-Hausdorff and interleaving distances between trees}
% \author{Elena Farahbkhsh Touli} \and \author{y}
%\date{}

%\begin{document}
\maketitle
%\linenumbers
\setcounter{page}{1}

\input{Intro1}

\vspace{-1mm}
\section{Problem Description and Background}
\label{Prob_Desc}
%{\color{blue}There are} 
 Consider a system with $(n+1)$  players  involved in a resource sharing game (RSG) and  let  $N = \{0, 1, 2, \cdots, n\}$  %{\color{blue}denotes} 
 denote the set of players  along with an adamant player represented by index $0$.  The  $n$ players (other than the adamant player)   are %{\color{red} symmetric (,i.e., players have same influence factors) and are}
 interested in forming coalitions, and these  are referred to as C-players. These players are willing to cooperate with each other if they can obtain higher individual share while the adamant player is not interested in cooperation. 

The utility of players is  proportional to their actions  which also includes a proportional cost. %and they incur a cost which is also proportional to their actions.
 Thus, when players choose respective actions $(a_0, a_1,  \cdots, a_n)$, the utility of player $i $ equals \vspace{-5mm}
%\begin{equation} \hspace{4mm}
%U_i = \frac{\lambda_ia_i}{\sum_{j=0}^n \lambda_ja_j} - \gamma a_i \; \forall \; i \; \in \; N \mbox{, where,  }
%\label{eq:Util_NC}
%\end{equation}
\begin{equation} \hspace{4mm}
\varphi_i = \frac{\lambda_i a_i}{\sum_{j=0}^n \lambda_j a_j} - \gamma a_i \; \forall \; i \; \in \; N \mbox{, where,  }
\label{eq:Util_NC}
\vspace{-2mm}
\end{equation}
\begin{tabular}{l  l}
  $\gamma$  \ represents the cost factor, \\
   $\lambda_i$  represents the influence factor of $i^{th}$ player, and,  \\
 $a_i $  represents the action of $i^{th}$ player with  \\  
  $a_i \; \in \; (0,\hat{a}) \; \text{for some} \;  n /\gamma  < \hat{a} < \infty $  which ensures the \\
  \ \, \, existence of a unique Nash Equilibrium  (\cite{dhounchak2019participate}). \hspace{5mm}
\end{tabular}
The first component of equation \eqref{eq:Util_NC} is the fraction of resource allocated to player $i$ and the other component represents the cost.
This resembles the utility of players in well known Kelly-mechanism for resource sharing and is relevant in various applications including communication networks (e.g., \cite{kelly1998rate,tun2019wireless} etc.), % which can model a variety of applications including 
online-auctions (\cite{koutsopoulos2010auction}), etc. 

In this paper, we consider the case with symmetric C-players, i.e.,  $\lambda_i = \lambda$ for all $i \in N_C$ where $N_C := \{1,\cdots,n\}$ is the set of C-players and $\lambda_0 = \eta \lambda$ with $\eta \in [0,\infty)$. Here, $\eta =0$ implies the absence of adamant player (considered in Section \ref{no_adamant_player}).

When the players choose their actions in a fully non-cooperative manner, i.e., when none of the  C-players are   interested in forming coalitions,  it results in a strategic form game with utilities as in \eqref{eq:Util_NC};
basically the rational and intelligent players choose their respective   actions to improve their own utility and 
the  utility derived by any player equals that  at Nash Equilibrium (NE){\footnote{NE is a well-known solution concept for non-cooperative games. For a strategic form game $\langle N,X,U \rangle$, a strategy profile $\underbar{x}^* = (x_1^*, \cdots, x_n^*)$ is called a NE of a strategic form game, if
		$
		U_i(x_i^*,x_{-i}^*) \geq U_i(x_i,x_{-i}^*) \, \forall \, x_i \in X_i,  \ \forall \, i \in N.
		$. This solution is stable against unilateral deviation.}}.

%{\color{blue}The  utility  in  \eqref{eq:Util_NC} is an adversarial extension of very well known Kelly-mechanism for optimal resource sharing, particularly studied for communication networks (e.g., \cite{kelly1998rate,tun2019wireless} etc)  which can model a variety of applications including online-auctions (\cite{koutsopoulos2010auction}), social networking, etc. } {\color{blue}We consider $\hat{a} > n / \gamma$, which ensures the existence of (unique) Nash Equilibrium (NE) (e.g., \cite{dhounchak2019participate}).}

  When  the $C$-players are looking   for opportunities to form coalitions and work together, a set/collection of coalitions   emerge at an appropriate equilibrium (details in  later sections);  say   ${\cal P} = \{S_0,S_1, \cdots, S_k\}$   represents the partition of $N$ into different coalitions where $S_0 \hspace{-1mm} = \hspace{-1mm} \{0\}$ denotes  the adamant player. Observe that a partition \( \mathcal{P}  \) is a set of coalitions such that \vspace{-2mm}
\begin{equation}
 \cup_{i=0}^k \; S_i = N \; \text{and} \; S_i \cap S_j = \emptyset, \mbox{  null set, } \, \forall \, i \neq j.
\label{eq:partition}
\end{equation}
The players in a \textit{coalition} $S_i$  choose their strategies together with an   aim to optimize  their social objective function (of their own coalition) and hence the utility of a coalition is given by:
%\vspace{-6mm} 
%{\small
\begin{equation}
\label{Eqn_Util_coaltiion}
 \hspace{-1mm} \varphi_{S_m} ({\bf a}_m, {\bf a}_{-m} ) = \frac{\lambda\sum_{l \in S_m} a_l }{\lambda_0a_0+\lambda\sum_{l=1}^n a_l} - \gamma \sum_{l \in S_m} a_l; \; m \geq 1
 \end{equation}
 \vspace{-2mm}
\begin{equation} 
\label{Eqn_Util_coaltiion_adamant} \varphi_{S_0} ({\bf a}_m, {\bf a}_{-m} ) = \frac{\lambda_0a_0 }{\lambda_0a_0+\lambda\sum_{l=1}^n a_l} - \gamma a_0, \text{where,} 
\vspace{-2mm}
\end{equation}
\begin{eqnarray}
 \ {\bf a}_m = \{a_i, i \in S_m\}, {\bf a}_{-m} = \{a_i, i \notin S_m\},  \forall  S_m \in {\cal P}, \nonumber
\end{eqnarray}which is the sum of their individual utilities.
The players will now try to derive maximum utility for their own coalition and hence  there would again be a non-cooperative game, %{\color{blue}between} 
 but now  among coalitions. Thus we have  a reduced RSG (one for every $\cal{P}$) with each coalition representing one (aggregate) player and  the utilities given by \eqref{Eqn_Util_coaltiion} and \eqref{Eqn_Util_coaltiion_adamant}; utility of  any coalition  equals that at the corresponding NE. % of the reduced/aggregate game.  
 This utility is divided among the members of the coalition using the well-known  {\it Shapley values} (computed within the coalition)  as described in Section \ref{sec_div_worth}.

 This is the problem setting and our aim is to study the  coalitions/partitions that  emerge out successfully (at an appropriate equilibrium), when the C-players (henceforth referred as players) seek opportunities to come together in a non-cooperative manner.
There is a brief initial study of this problem in   \cite{dhounchak2019participate}, for the special case when players only form grand coalition, i.e., when ${\cal P} = \{ \{0\}, \{1,  \cdots, n\}\}$. For this case, it has been shown that: 

i)  The utility of grand coalition at CNE (Cooperative NE) is higher than the sum of individual utilities of players at the unique NCNE (Non-Cooperative NE) for majority of the scenarios. The paper also provides example scenarios for the %{\color{blue}other} 
case where the sum of utilities at NCNE is larger.% than the  utility of grand coalition of   C-players at CNE.

ii) Moreover, Shapley value does not always share this utility  in a fair manner;   the sum of utilities might be larger, but  the shares derived via Shapley value by some  players  is smaller (especially ones with higher influence factors).   This aspect is not further investigated in the current paper.

%iii) Also, there were some scenarios where the sum of utilities at NCNE is larger than the  utility of grand coalition of   C-players at CNE. 
%{\color{red} The result is only for asymmetirc C- players. For symmetric players when $n=2$, it is true.
%}

The above study lead to new questions: a) can the players derive even  better utilities if they form strict sub-coalitions instead of grand coalition; b)  when  is it  beneficial for the players to cooperate; c) how stable are these resultant coalitions  (e.g., against unilateral deviation). One can have more questions with asymmetric C-players and we wish to investigate this in future (we already have some initial results\footnote{With one asymmetric player and no adamant player, we found that grand coalition can also emerge at NE, even for large $n$.}). 

   {\it We  build an appropriate non-cooperative framework to study these aspects. We also consider  solutions that optimize social objective function and {\it derive the Price of Anarchy.}} 
 
 \section{Adamant Coalition Formation Games}

 \label{ACFG}
We use non-cooperative framework to study this coalition formation  game (CFG) as in \cite{Nevrekar2015ATO}. 
 For each C-player, i.e., for $i \in N_C$, strategy $x_i$  is defined as the  set of players with whom player $i$ wants to form coalitions, i.e., $x_i \subseteq N_C$ and the corresponding strategy set $X_i$  is defined as:\vspace{-2mm}
$$
X_i = \{x_i : i \in x_i \mbox{ and } x_i \subseteq N_C \}.
\vspace{-2mm}
$$
To construct a  strategic form game we need to define the utility of all players for any given strategy profile, i.e., for any $\underbar{x} = (x_1,x_2,\cdots,x_n) \mbox{ with } x_i \in X_i \mbox{ for each }i \in N_C$.

As a first step, one needs to define appropriate partition(s) of coalitions (referred as $\mathcal{P(\underbar{x})}$,  and made up of subsets of  $N$)  that can result for any   given  strategy profile $\underbar{x}$. 

\vspace{-1mm}
\subsection{Partition for a given strategy profile $\underbar{x}$}
\label{sec_partition_given_strategy_profile}
%\vspace{-1mm} 
%We begin with few definitions. 
 We say a {\it partition  $\mathcal{P}'$ is (strictly) better than partition $\mathcal{P}$, represented by the symbol $ \mathcal{P}' \prec \mathcal{P}$,}  if every coalition of the latter is a subset of a coalition of the former (with at least one of them being a strict subset), i.e.,   if
    \vspace{-2mm}
  \begin{eqnarray}
    \mathcal{P}' \hspace{-1mm} \ne \hspace{-0.5mm} \mathcal{P}  \mbox{, and,       for all } S \hspace{-0.5mm} \in \hspace{-0.5mm} \mathcal{P}    \
   \exists \   S' \hspace{-1mm} \in \mathcal{P}'  \mbox{ such that } S \subset S'.  \hspace{-2mm} 
   \label{Eqn_Max_Coalition_rule_0}
   \vspace{-4mm}
\end{eqnarray}
\vspace{-6mm}

 Note that the size   (number of coalitions)  of the better partition is strictly smaller than that of the other; in other words, there exists at least two coalitions $S_1, S_2  \in {\cal P}$ such that 
$S_1 \cup S_2 \subset S$ for some $S \in {\cal P}'$.

{\it  Partition $\mathcal{P}(\underbar{x})$  formed  by $\underbar{x}$:}
We say ${\underbar x} \to  \mathcal{P} (\underbar{x})$, if  it satisfies the following two conditions as in \cite{Nevrekar2015ATO}:\\
i) {\it respects the preferences,} 
a coalition $S$ is an element of partition $\mathcal{P(\underbar{x})}$, i.e.,  $S \in \mathcal{P(\underbar{x})}$,  if  it satisfies: 
\vspace{-2mm} 
\begin{eqnarray}
 i \in x_j  \mbox{ and }  j \in x_i  \mbox{   for all }  i, j  \in S;  \mbox{ and, }  \label{Eqn_Coalition_rule}  
 \end{eqnarray} 
ii) {\it minimal partition,} there exists no other (see \eqref{Eqn_Max_Coalition_rule_0})
\vspace{-2mm} 
\begin{eqnarray}
\mbox{(better) partition  $\mathcal{P}' $ formed by } \underbar{x}  \mbox{, such that  } \mathcal{P}' \prec \mathcal{P}.    \label{Eqn_Max_Coalition_rule}
\end{eqnarray}
 
%    Further  the  eventually resulting coalitions   should not be a subset of other coalitions that potentially satisfy the requirement given by equation \eqref{Eqn_Coalition_rule}. 
Hence, a \emph{partition} formed by $\underbar{x}$ is a (minimal) subset of $2^N := \{S: S \subset N\}$  such that   \eqref{eq:partition} and  \eqref{Eqn_Max_Coalition_rule}  are satisfied and all its coalitions satisfy  \eqref{Eqn_Coalition_rule} with $\underbar{x}$. Using these rules, we may obtain multiple partitions for some strategy profiles (examples  in Tables \ref{tab:n=3} and \ref{tab:n=4}). 

To summarize, if 
$\underbar{x} = (x_1, \cdots, x_n)$ is the strategy profile, let $n(\underbar{x})$ represent the number of possible partitions   corresponding to $\underbar{x}$, and  let  the partitions formed be  represented by the following:
\vspace{-2mm}
$$
{\cal P}^1 (\underbar{x}), \, {\cal P}^2 (\underbar{x}) \,  \cdots \, {\cal P}^{n\small{(\underbar{x})}}  (\underbar{x}).
 \vspace{-2mm}
$$
We now define the utilities derived by (all) the coalitions and then the individual players. We begin with   $n(\underbar{x}) = 1$.
\subsection{Utilities of coalitions in a given partition}
\label{sec_util_coal}
Let $\mathcal{P(\underbar{x})}$ = $\{S_0,S_1,   \cdots, S_k \}$ be a  partition of $N$  with $k$ coalitions of C-players,   corresponding to $\underbar{x}$. We now  aim   to find the utility of %adversary, $U_{S_0}$ and  that of 
coalitions in $\mathcal{P}(\underbar{x})$, represented by  $\varphi^*_{S_m}(\mathcal{P})$ for all   $ m \in \{0,1,2, \cdots, k\}$. We will see that  these utilities depend upon the strength of the adamant player, via $\eta := \lambda_0 / \lambda$, the relative ratio of the influence factors (recall $S_0 = \{0\}$ is the coalition with  only adamant player). 
%  An adversary is considered as a coalition with  single player.

As already mentioned, \Cmnt{the utility of any coalition depends upon the joint actions of players of each coalition    (see     (\ref{Eqn_Util_coaltiion})); basically each coalition  can be seen as a single player; all the players in any  coalition    strive together to achieve the best for their own coalition.  So,} the resource sharing game (RSG) is now reduced to a \textit{(k+1)-(aggregate) player non-cooperative strategic form game} which is given by the tuple, 
\begin{eqnarray}
\label{Eqn_reduced_RSG}
 \bigg \langle \{0, 1, \cdots, k\}, \{[0,\hat{a}]^{|S_0|} \times \cdots\times [0,\hat{a}]^{|S_m|}\},\mathbf{\varphi} \bigg \rangle,
\end{eqnarray}
  where $|S_m|$ represents the cardinality of coalition $S_m$ and $\mathbf{\varphi}$ = $\{\varphi_{S_0}, \varphi_{S_1}, \cdots \varphi_{S_k} \}$, the vector of utilities is given by  \eqref{Eqn_Util_coaltiion} and \eqref{Eqn_Util_coaltiion_adamant}.
%
%This leads to k+1 player game. 
%
This kind of a game is analysed in \cite[Lemma 2]{dhounchak2019participate} for the  special case with grand coalition (GC) of C-players. \Cmnt{According to this Lemma (with symmetric C-players),  at an NE\footnote{ There are multiple NE, but the utility of GC is the same at all of them.},   
all C-players except for one remain silent; this silencing reduces the  interference\footnote{the non-zero actions (at NE)  of the players increases the denominator in \eqref{Eqn_Util_coaltiion} at NE, hence the utilities are reduced. } and the sum utility of the GC improves.  However, this silencing   reduces interferences  not only  to the members in its own  coalition, but also to  the players outside the coalition;  this will have a major impact on our CFG (as will be seen soon).  }
Since, we consider all possible exhaustive and disjoint collection of players, i.e., all possible partitions (corresponding to various coalition suggestive strategy profiles), we  extend the above  result to a general partition in the following:
\begin{theorem} \textbf{[Utilities of coalitions]}
	\label{thm:Thm1}
	  The game \eqref{Eqn_reduced_RSG} can have multiple NE, but the utilities at NE are unique and are given by (for any $1\leq m \leq k$),
	
	\vspace{-4mm}

		{\small \begin{eqnarray}
		\label{Eqn_USm}
		\varphi^*_{S_m}(\mathcal{P}) & = & \frac{\lambda^2}{  (\lambda  +   k  \lambda_0 )^2} \mathbbm{1}_B 
		+ \frac{1}{ k^2   }  (1 - \mathbbm{1}_B) \nonumber \\ &=&\frac{1}{  (1  +   k  \eta )^2} \mathbbm{1}_B 
		+ \frac{1}{ k^2   }  (1 - \mathbbm{1}_B),
		\end{eqnarray}}
		 with indicator  $\mathbbm{1}_B := \mathbbm{1}_{\eta > \frac{k-1}{k}}$, $k =  | {\cal P} | -1$,  $\eta  := \lambda_0/\lambda$  and, 
		 {\small \begin{eqnarray}
		 	\label{NE_util_adamant}
	\varphi^*_{S_0}(\mathcal{P}) = \Big(\frac{(1-k)\lambda+k\lambda_0}{  \lambda  +  k  \lambda_0 }\Big)^2 \mathbbm{1}_B = \Big(\frac{(1-k)+k\eta}{  1  +  k  \eta }\Big)^2 \mathbbm{1}_B.   
		\end{eqnarray}} Further the optimal actions at any NE  satisfy:
	{\small {\begin{equation}
	{\bar a}^*_m := \hspace{-1mm} \sum_{j \in S_m} a_j^* =    \frac{k\lambda\lambda_0}{\gamma (\lambda + k \lambda_0)^2} \mbox{ \normalsize and, } {\bar a}^*_0 = \frac{k\lambda((1-k)\lambda+k\lambda_0)}{\gamma (\lambda + k \lambda_0)^2}.
	\label{Eqn_alstar}
	\end{equation}}}
		 	
	\textbf{Proof: }The proof is almost similar to the one in \cite{dhounchak2019participate} and is available in Appendix A.
\eop	
	
\end{theorem}

 \Cmnt{We basically derive the utility of adamant player and that of the    coalitions  of the partition at a NE of this %which are basically the utilities at NE of the
	 RSG.   The proof is almost similar to the one in \cite{dhounchak2019participate} and is available in Appendix A; we again have multiple NE but the utilities of the coalitions at all the NEs are the same. We reproduce these utilities for any given partition for ease of reference;
for any coalition $S_m \in \mathcal{P}$ (with  $m \geq 1$):

\vspace{-4mm}
{\small
\begin{eqnarray}
\label{Eqn_USm}
 \varphi^*_{S_m}  =  \frac{\lambda^2}{  (\lambda  +   ( | {\cal P} | -1)  \lambda_0 )^2} \mathbbm{1}_B 
 + \frac{1}{ (| {\cal P} |-1)^2   }  (1 - \mathbbm{1}_B), \\
 \nonumber  
  \mbox{ \normalsize with indicator } \mathbbm{1}_B := \mathbbm{1}_{\eta > \frac{|\mathcal{P}|-2}{|\mathcal{P}|-1}}  \mbox{ \normalsize and  }  \eta  := \lambda_0/\lambda.  
\end{eqnarray}}}

From \eqref{Eqn_alstar},  some or all of the players in a coalition can choose actions such that the sum of these actions equal corresponding $\bar{a}_m^*$; all such actions consitute NE; hence multiple NE exist. However, the utilities of coalitions are uniquely defined by  \eqref{Eqn_USm}.

\noindent
{\bf Significant Adamant Player:}
  The adamant player gets non-zero utility at NE   when  $\mathbbm{1}_B = 1$,  i.e.,  when $\eta > 1 - 1/ k$,  we then say the adamant player is significant otherwise, it is insignificant.  However, it is always significant when grand coalition is formed, i.e., when $k=1$ (see equation \eqref{NE_util_adamant}).  This condition will play an important role in our CFG. 

To summarize the utilities of any coalition of any given partition $\mathcal{P}$  are given by \eqref{Eqn_USm} and \eqref{NE_util_adamant}, which  are the utilities at NE of the reduced RSG with coalitions  as the players.

\subsection{Division of worth  within a coalition}
\label{sec_div_worth}
The next step is to divide the worth of a coalition among its members  using Shapley value confined to each coalition as in \cite{aumann1974cooperative}.
 For  symmetric players, the utility of a coalition gets divided equally among its members because of equal influence factors. Hence from \eqref{Eqn_USm}, the utility of  player $i$  under  partition 
${\cal P} $ is given by (if $i \in S_m$):
%\vspace{-1mm}
\begin{equation}
\label{Eqn_USm_player}
\varphi^*_{i} ({\cal P})   = \frac{   \mathbbm{1}_B  }{  |S_m|  \left (1 +  k   \eta  \right )^2}
+ \frac{1- \mathbbm{1}_B }{ k^2  |S_m|  } \mbox{ with } k=|{\cal P}|-1.
\end{equation}
\vspace{-1mm}

\vspace{-4mm}
\subsection{Utility of a player}
\label{sec_util_player}
 We define the utility of a player, say $i$ as the minimum utility among all the possible partitions, i.e.,  (see \eqref{Eqn_USm_player})
  %\vspace{-1mm}
	\begin{equation}
	\label{min_util_mult_partition}
	U_i(\underbar{x})= \min_{\mathcal{P}({\small \underbar{x}})} \varphi^*_i (\mathcal{P}(\underbar{x}) ).
	\vspace{-1mm}
	\end{equation} 
	This definition ensures minimum guaranteed utility to each player for the given strategy profile $\underbar{x}$ and is similar to the \textit{security value} used in game theory (\cite{narahari2014game}). Basically, when a strategy profile (recall it represents the coalition formation interests of all the players) can lead to multiple partitions, %say $n(\underbar{x})=n'$ (recall $n(\underbar{x})$ is the number of partitions resulting from a strategy profile $\underbar{x}$)
	 the eventual partition formed may depend on some further negotiations. %In such a scenario, each player derives one of the $n'$ possible values of utility corresponding to the partition agreed upon.
	  Hence it is best to define the utility of each player as the worst possible  utility.%  (,i.e., what they can get for sure).   
	%when any one of the $n'$ partitions is agreed upon. 

\vspace{-1mm}
\subsection{Coalition Formation Game: Ingredients}
\label{sec_CFG_ingre}
\label{sec:CFG}
We now have a non-cooperative CFG with,  i) $N_C  $ as the set of players;
ii) $
X_i$ is the strategy set of player $i $; and % is 
%$
%X_i = \{x_i : i \in x_i \,; x_i \subseteq N_C \}; 
%$
iii) Utilities  of   players, $\{U_i ({\underbar{x} }) \}_{i,{\small{\underbar{x}}}}$ %{\color{red} $=\{ \varphi_i(\mathcal{P}({\small{}\underbar{x}))\}_i$}
   given by \eqref{min_util_mult_partition}. Recall  these utilities are defined via their Shapley value corresponding to the coalition that they belonged (based on their and others strategies), the worth of which is computed using NE of the reduced RSG. %between various members (i.e., coalitions) of the partition so formed.}{.}

We study this game and consider two types of solution concepts: NE and Social Optima and also discuss the price of anarchy in the coming sections.

%\vspace{-1mm}
\section{Initial Analysis and Solutions}
\label{init_analysis}
%We begin with analysis of symmetric game, i.e., where C-players have same influence factor $\lambda$.
%We will observe that this analysis splits into two types depending upon the number of C-players. 
In this section we consider some partition-wise analysis which later leads to the analysis of Nash equilibrium.  We also define the solution concepts used in this paper. 
%We begin this section with few definitions.

 \vspace{-1mm}
\subsection{Partition resulting from a unilateral deviation} 
\label{sec_partition_ud}
Recall a  {\it strategy profile}  $\underbar{x}$ {\it leads to  partition $\mathcal{P}$},  represented by $\underbar{x} \to \mathcal{P}$,  if    $\mathcal{P}$  results from   $\underbar{x}$ as explained in section \ref{sec_partition_given_strategy_profile}, i.e., if  it satisfies \eqref{Eqn_Coalition_rule} with $\underbar{x}$, \eqref{eq:partition} and \eqref{Eqn_Max_Coalition_rule}. 
We say,     $\underbar{x}$ {\it leads to unique partition $\mathcal{P}$},  represented by $\underbar{x} \to !\mathcal{P}$,  further,  if  $\mathcal{P}$ is unique such partition, i.e., if $n(\underbar{x}) = 1$.

Consider  any  partition $\mathcal{P} = \{S_0, \cdots,S_k\}$ %and  without loss of generality\footnote{because of symmetric players} (wlog),  say 
 %
% \vspace{-4mm}
% {\small \begin{eqnarray*}
% %
% S_1 = \{1, \cdots, m_1\}, \  S_i = \{m_{i-1}+1, \cdots, m_i\} \,  \forall \, 1 < i \le k
% \end{eqnarray*}}for some $m_1,m_2,\cdots,m_k$,
% 
  and say $ \underbar{x} \to !\mathcal{P}.$
%{\color{blue}Consider}
Now consider a  unilateral deviation    of  player  $i$,  from  $x_i$  to  $\{i\}$ (strategy of being alone)  in   $ \underbar{x}$ and say $i \in S_l$.   % and   say $i \in S_l$ of partition $ {\cal P}$, for some $l \leq k$. 
 Then, the following lemma shows that the new strategy profile ($\underbar{x}'$) also leads to a unique partition with $S_l$ coalition getting split into two; $\{i\}$ and $S_l/\{i\}$ (the rest as one sub-coalition):
\begin{lemma}
\label{Lem_partiton_uni_dev}
Consider a strategy profile  $\underbar{x} \to !\mathcal{P}$, where  $i \in S_l$. 
Let  $\underbar{x}' = (\{i\},  \underbar{x}_{-i})$ be the strategy obtained by the above unilateral deviation, then    $  \underbar{x}'  \to  !{\cal P}_{-i}$, where: 
 \vspace{-1mm}
  \begin{equation*}
 \mathcal{P}_{-i} :=   \{S_0,S_1, \cdots, \  S_{l-1}, \{i\},  \  S_{l} \backslash   \{i\},   S_{l+1}, \cdots, S_k\} .
 \vspace{-2mm}
  \end{equation*}%{}\color{blue}We
  %also have uniqueness, i.e., $  \underbar{x}'  \to  !{\cal P}_{-i}$.}
    %Further, such a partition ${\cal P}_{-i}$ is unique, i.e.,  $  \underbar{x}'  \to  !{\cal P}_{-i}$.
\end{lemma}

\textbf{Proof :} The proof is in Appendix B.  \eop
% \TR{Appendix B}{\cite[Appendix B]{TR}}. \eop 

%{\color{blue}Here the $S_l$ coalition gets split into two sub-coalitions $\{i\}$ ($i$   alone) and  $S_{l} \backslash   \{i\}$ (the rest as one sub-coalition).} 
 %
We call partition  $\mathcal{P}_{-i} $ of the above Lemma  as the {\it $i$-unilateral deviation  partition,  $i$-u.d.p.,}  of  the pair  $(\underbar{x}, \mathcal{P})$.

\subsection{Weak Partition}
\label{sec_weak}
 A \emph{partition is defined to be weak} if  for all $ \underbar{x} \to !\mathcal{P}$, there exists a  player $i$  which gets strictly better utility  at  its  $i$-u.d.p, i.e.,   if ($\mathcal{P}_{-i}$ defined  in Lemma \ref{Lem_partiton_uni_dev})
 \vspace{-2mm}
 $$U_i(\mathcal{P}_{-i})>U_i(\mathcal{P}). \vspace{-2mm}$$
  % We say partition ${\cal P}$  is \emph{strong} if no player can improve from it by deviating unilaterally.

With the above definitions in place, we have the following result for characterizing the weak partitions:
\begin{lemma}
\label{Lemma_Partition_weak}
Consider a partition $\mathcal{P}$ with {\small$|\mathcal{P}| = (k+1)$}.    Let  $m^* := \max_{S_i \in \mathcal{P}} |S_i|$, be the  size of the largest coalition. If $m^* >  (k+1)^2/ k^2 $, then   $\mathcal{P}$ is \emph{weak}. 
\end{lemma}
{\bf Proof :}
The proof is   in Appendix B. \eop 
%\TR{\cite[Appendix B]{TR}.}{Appendix B.}

\noindent
\textbf{Remark: }Say for all the strategy profiles $\underbar{x}$ leading to  $\mathcal{P}$ it is the unique such one (i.e., $\underbar{x} \to !\mathcal{P}$). Further, if it satisfies the above conditions, it  cannot be a partition at NE. However, if there is a strategy profile leading to multiple partitions with one of them being $\mathcal{P}$, then $\mathcal{P}$ can  still emerge at a NE. We will  investigate these aspects in the immediate following.

\subsection{Nash Equilibrium }
\label{sec_NE}
To study the CFG (see section \ref{sec:CFG}), we again consider the solution concept  \emph{Nash Equilibrium (NE)} \cite{narahari2014game} (recall this solution ensures that no player can get better on unilateral deviation).  The NE is now in terms of coalition suggestive strategy profile, but one might be more interested in
   \textit{NE-partitions}.
 Lemma \ref{Lemma_Partition_weak} characterizes  weak partitions, and one may think weak partitions  cannot   result from a NE. However, as discussed before, if a weak partition is one amongst the multiple partitions emerging from  a NE, then a weak partition can also be a NE-partition. Thus we have:
\begin{lemma}{\bf [NE $\not \to$ Weak Partition]}
\label{Lemma_weak_NE}
Assume that the game does not have multiple partitions at NE. %{\color{red} Assume that the strategy profile leading to multiple partitions cannot be a NE.}
 Then, if a partition ${\cal P}$  is weak,  it cannot be a NE-partition. %result from a NE  of the coalition formation game.  

\end{lemma}
\textbf{Proof :}
The proof is straightforward.
\eop

If for a  given set of parameters, it is known a priori that none of the NE   lead to multiple partitions, then by the above Lemma,     a weak partition can't emerge from a NE. We will then concentrate on  partitions that are not weak. 
We will use these intermediate results to derive the NE.  Before we proceed with this we discuss the relevant  social objective function.

\subsection{Social Optima}
\label{sec_SO}
In this paper we are primarily studying the CFGs, in which the players choose their partners in a non-cooperative manner; basically the players are interested in coalition formation, so as to  improve their own objective function (selfishly)  and one requires a solution which is stable against unilateral deviations. But if instead the players attempt to optimize a social/utilitarian objective (sum of utilities of all the players),  they would have achieved much better utilities; this aspect   is well understood in literature (\cite{johari2004efficiency} and references therein) and we study the same in our context.   
 A \emph{utilitarian solution}, referred to as SO (social optimizer),  is any  strategy profile  $\underbar{x}_S^
 *$ that maximizes: 
 \vspace{-1mm}
$$
\sum_{i \in N_C}U_i(\underbar{x}_S^*) = \max_{{\small{\underbar{x}}} } \sum_{i \in N_C}U_i (\underbar{x})  . % \; \forall \, i=1,2,\cdots,n.
\vspace{-1mm}
$$
In \cite{dhounchak2019participate} authors illustrated that the sum utility of the  C-players improve significantly, when all players come together to  form a grand coalition (as $n$ increases). However we will see in this paper that for $n > 4$, the only  NE-partition is ALC (all alone). Because of the
 selfish nature of the players, the efficiency of a system degrades and the utility received by players at NE is lower than  that at SO.
%This can be measured by \emph{Price of Anarchy}.
We study  this loss  using the well known concept, \emph{Price of Anarchy}.

One might be interested in the NE or SO, basically the strategies that represent the solutions.  However in our context,  the {\it more interesting entities are the partitions at various equilibrium/optimal solutions;}   we are interested in  NE-partitions  and  the SO-partitions.
%
% Also, for this case instead of summing up the individual utilities of a player corresponding to a strategy profile $\underbar{x}$ we can simply sum the utilities of coalitions corresponding to a partition emerging from strategy profile $\underbar{x}$. This is because the coalitional utilities are divided among the members of coalition according to their influence factor, thus leading to their individual utilities.
When one directly optimize\footnote{One can think of such an optimization,  as   the players are working together now. } using partitions;  it is easy to see that the  SO-partition, 
${\cal P}^*_S$   satisfies the following:
\vspace{-1mm}
$$
U_{SO}^* := \sum_{S_i \in \mathcal{P}^*_S;i\neq0}U_{S_i}( \mathcal{P}^*_S) = \max_{\mathcal{P}'} \sum_{S_i' \in \mathcal{P}'; i \neq 0}U_{{S_i}'}( \mathcal{P}'),  % \; \forall \, i = 1,2,\cdots,n
\vspace{-2mm}
$$
where $\mathcal{P}'$ includes all possible partitions %{\color{red} corresponding to a given strategy profile}
 (see \eqref{Eqn_USm}-\eqref{Eqn_USm_player}).  

\subsubsection*{\bf Some more notations}
{\it Let ${\cal P}_k$ represent any partition with $k$  coalitions of C-players}, i.e.,  $| {\cal P}_k | = 1+k$. 
Let {\it  (All aLone Coalitions) ALC}$:={\cal P}_n = \{ \{0\}, \{1\}, \{2\}, \cdots, \{n\}\}$ (all players are alone).   
The  strategy $x_i = \{i\}$  is the  ALC strategy for any $i$, and 
{\it GC (Grand Coalition}) partition implies partition $\{ \{0\},  N_C\}$, while  GC strategy implies  $x_i = N_C$.
%, and the corresponding strategy profile  of all C-players is also referred to as GC strategy profile.   

\textbf{Two groups of partitions: }
As seen in \eqref{NE_util_adamant},  at some equilibrium the {\it adamant player  becomes insignificant,} i.e., gets 0 utility.  
We distinguish  these  equilibrium partitions  from the others using superscript $^o$.  Thus, for example,  ALC is the NE-partition if adamant player is significant at that NE, otherwise, ALC$^o$ is the NE-partition. 

%
%It is appropriate to exclude $S_0 = \{0\}$ (coalition of adversary) from such partitions as this coalition does not alter any thing related to that equilibrium; 
%such partitions  are represented with super script $^*$ to distinguish them from the other group of partitions.  For example,  ALC includes adversary, while ALC$^* =  \{  \{1\}, \{2\}, \cdots, \{n\}\}$; note that $| {\cal P}_k | = 1+k$ while $| {\cal P}^*_k | = k.$

With the above notations in place, we have the following result completely characterizing the  SO-partitions:
\begin{lemma}{\bf [SO-partitions]}
\label{Lemma_SO_Partition}
i) When $\eta \ge 0.707$ or when $\eta \le 0.414$, then GC is the   SO-Partition.\\
ii)  When $0.414 \le \eta \le  0.5$,  any  ${\cal P}_2^o$  is the SO-partition.  \\
iii)  Any  ${\cal P}_2$  is a  SO-partition for  rest, {\small($ 0.5 < \eta \le 0.707$)}.  
\end{lemma}
\textbf{Proof :}
The proof is   in 
 Appendix B.
   \eop

\subsection{Price of Anarchy and SO-partition}
\label{POA}
Price of Anarchy ($P_{oA}$) is defined as the ratio between the sum utilities at  `social optima' and the sum utilities at the  `worst Nash Equilibrium', i.e., 
\begin{eqnarray*}
P_{oA}  &\hspace{-1mm}=\hspace{-1mm}& \frac{\max_{\mathcal{P}} \sum_{S_i \in \mathcal{P}; i \neq 0} U_{S_i}  }{ \min_{\mathcal{P}^*} \sum_{S_i \in \mathcal{P}^*;i \neq 0} U_{S_i}  }  = 
\frac{  U_{SO}^*  }{ \min_{\mathcal{P}^*} \sum_{S_i \in \mathcal{P}^*;i \neq 0} U_{S_i}  },
%=? \frac{ \sum_i u_i (AL)  }{  \sum_i u_i (GC)  }  \\
\end{eqnarray*}
where $\mathcal{P}^*$ is any NE-partition.

 \subsection*{ALC/ALC$^o$  is always an NE-partition }
When all others choose to be alone, i.e., if $x_i = \{i\}$ for all $i \ne j$, then it is clear that the best response of $j$  includes $x_j = \{j\}$. This is true for any $j$.
This leads to an NE.  
% Hence  ALC   is always an NE-partition. 
  From \eqref{NE_util_adamant}, the adamant player becomes insignificant at ALC when $\eta \le 1 - 1/n$, then the NE-partition is ALC$^o$, otherwise ALC is the NE-partition.

\section{Large  number of players, $n > 4$}
For the case with $n>4$, we have the following two results using Lemma \ref{Lemma_Partition_weak} (proofs in Appendix A):
\begin{corollary} \textbf{[Weak Partitions]}
All partitions other than ALC/ALC$^o$ are \textit{weak}. \eop
\label{corollary_ALC_weak}
\end{corollary} 
\Cmnt{{\color{red} Lemma  \ref{Lemma_weak_NE}  characterizes that NE-partitions can't be weak partitions, if the game does not have multiple partitions at NE; hence we first identify the weak partitions when $n>4$. The following corollary identifies the weak partitions when $n>4$ (proof in Appendix A).
	\begin{corollary}
		When $n>4$, all partitions other than ALC/ALC$^o$ are \textit{weak}. \eop
		\label{unique_weak}
	\end{corollary}}
\label{large_no}
{\color{blue}Lemma  \ref{Lemma_weak_NE}  characterizes that NE-partitions can't be weak partitions, if the game does not have multiple partitions at NE; %this provides a way to identify the NE-partitions. 
we first show that one can't have multiple partitions at NE, when $n > 4$ (proof in Appendix A).} % \TR{Appendix A}{\cite[Appendix A]{TR}}). }
{\color{red}Next we show that one can't have multiple partitions at NE, when $n > 4$ (proof in Appendix A).}}
\begin{theorem}{\bf [No Multiple Partitions at NE]}
\label{Thm_No_MPs_for_grt_n}
Any strategy profile leading to multiple partitions cannot  be an NE.  \eop %in symmetric players case, with adversary.
\end{theorem}
 In view of the above two results and Lemma \ref{Lemma_weak_NE}, only ALC/ALC$^{o}$ is the NE-partition. Further using \eqref{NE_util_adamant}, we have:   
%{\color{red} The above result when combined with Lemma  \ref{Lemma_weak_NE} helps us in reducing the strategy space to the strategy profiles leading to a unique partition which is not \textit{weak} and can be characterised using Lemma \ref{Lemma_Partition_weak}. In view of this it is easy to identify all the NE-partitions, which is precisely undertaken in the following corollary (proof in \TR{Appendix A}{\cite[Appendix A]{TR}}).}
\begin{corollary}
	\label{corollary_ALC_NE}
There is a unique NE-partition. 
%The only NE- partition, with number of agents larger than 4,   is ALC,  the partition in which all remain alone.  
ALC is the  NE-partition if $\eta > (n- 1)/n$, else 
ALC$^o$ is the NE-partition.  \eop
\end{corollary}

\textbf{Remarks:}  
In \cite{dhounchak2019participate}, authors defined BoC (benefit of cooperation) as the normalized improvement in sum of utilities that the players achieve at GC in comparison with that achieved when they  compete alone.  They showed that BoC increases significantly as $n$ increases (\cite[Lemma 3]{dhounchak2019participate}). 
Despite the fact that  BoC is large for large $n$, by the above Corollary  we have that  players prefer to remain alone at NE. Thus the  price paid for anarchy ($P_{oA}$)  can be significantly high.

\vspace{-1mm}
\subsection{ Price of Anarchy}
%Since, players are forming coalitions in a non-cooperative manner and hence their selfishness leads to lower utilities when they could have got better. This loss of players because of their selfish nature is known as \emph{Price of Anarchy}.

From Corollary \ref{corollary_ALC_NE}, 
we have ALC/ALC$^o$ as the only NE-partition and  from   Lemma \ref{Lemma_SO_Partition},  GC is the SO-Partition  when  $\eta \ge  0.707$. Hence $P_{oA}$ equals (see \eqref{Eqn_USm}): 

\vspace{-4mm}
%{\small
\begin{eqnarray*}
P_{oA} \hspace{-2mm} &=& \hspace{-2mm} \frac{\frac{1}{(1+\eta)^2}}  {\frac{n}{(1+n\eta)^2} }
\ = \  \frac{(1+n\eta)^2}{n(1+\eta)^2} \mbox{ when }  \eta \geq 0.707. \Cmnt{\mbox{ and, } \\
P_{oA} \hspace{-2mm}
&=& \hspace{-2mm} \frac{ n }{(1+\eta)^2} \mbox{ when }  \eta \le  0.414 \text{ or }   0.707 \le \eta \leq \frac{n-1}{n}.}
\end{eqnarray*}
We compute $P_{oA}$ for the remaining  cases  in a similar way and the results are  in Table \ref{tab_PoA_large}.
Clearly as $n \to \infty$, $P_{oA}$   grows like   $n$, i.e., $P_{oA} = O(n)$;   this is another instance of strategic behaviour where the players pay high price for being strategic. % hence confirming our initial  conjectures.   

\subfile{Social_optima_n_small}

\vspace{-2mm}
\subsection*{$P_{oA}$ and  Observations}
The $P_{oA}$ for smaller $n$ is computed in the Tables \ref{tab:my-table2}, \ref{tab:my-table1} and \ref{tab:my-table}, and the overall observations related to $P_{oA}$ are:

i) From Table \ref{tab_PoA_large}, as the number of players increases the  $P_{oA}$ also increases, and, $P_{oA} = O(n)$ when $n \to \infty$;  

ii) For  any $n$ as the adamant player grows strong (as $\eta \to \infty $),  the $P_{oA} \uparrow  n$ (see Tables \ref{tab_PoA_large}, \ref{tab:my-table2}, \ref{tab:my-table1}, \ref{tab:my-table});   and

iii) Similarly  when the adamant player becomes weak ($\eta \to 0 $),   the $P_{oA}$ again increases to  $ n$.  
\section{Without Adamant Player}
\label{no_adamant_player}
We now consider the same model as in  previous sections, but without adamant player. %  with the only difference being the absence of adversary.  
Majority of the analysis goes through as in previous cases, we will only mention the differences. The utility of any  partition
  $\mathcal{P}^o = \{S_1,\cdots,S_k\}$ and that of the individual players, using Theorem \ref{thm:Thm1}   and Shapley value  simplify to:  

\vspace{-5mm}
%{\small
	\begin{eqnarray}
\label{Eqn_USm_without_adv}
\label{Eqn_Ui_without_adv}
U_{S_m}   =   \frac{1}{ | {\cal P}^o|^2   } \forall m,  \mbox{ and }   U_i  =   \frac{1}{ | {\cal P}^o|^2  |S_m| }  \mbox{ if } i \in S_m.
\end{eqnarray}
These utilities are exactly the  same as those in the previous model with insignificant   adamant player, except for GC.

The results for the case with $n >4$ are exactly the same because of the following:
i) Lemmas \ref{Lem_partiton_uni_dev} and \ref{Lemma_weak_NE} are independent of adamant player;
ii) Theorem \ref{Thm_No_MPs_for_grt_n}  is also applicable,  since  only  ${\underline x}_G := \{N_C,  \cdots, \ N_C\}$ leads to   $GC$, and that too\footnote{Since in the previous case (in the presence of adamant player), adamant player was always significant; but this is not true here. } ${\underline x}_G \to! GC$; and 
iii) The proof of  Lemma \ref{Lemma_Partition_weak}  can easily be adapted.

\subsubsection*{Smaller n}
\TR{ When $n=2$, the only possible partitions for this case are: GC$^o$ and ALC$^o$. 
Consider best response of a player 1 against strategy $\{1,2\}$ of player 2. From \eqref{Eqn_Ui_without_adv}, the utility of player 1 if it chooses $\{1,2\}$ (GC$^o$ is formed) is
\begin{eqnarray*}
\frac{1}{2} & > & \frac{1}{2^2}
\end{eqnarray*}
the utility of the same player if it chooses to be alone.  In a similar way one can compute NE for $n = 3$ and $ n = 4$} {One can compute NE for all these cases as before,} and the results are in Table \ref{tab_without_adv}.  %Thus all the NE-partitions are the same as those with adversary with  $\eta \le 0.414.$
 In similar way the SO-partition is GC$^o$  (proof in  Appendix B):

\begin{lemma}
\label{Lemma_SO_no_adv}
GC$^o$ is  the  SO-partition in the absence of adamant player, for all $n$. \eop
\end{lemma}

%\vspace{-2mm}
\begin{table}[h]
\centering
\begin{tabular}{|c|c|c|c|}
\hline
$n$  &  $\mathcal{P}^o$ at NE&  $\mathcal{P}^o$ at SO & $P_{oA}$ \\ \hline
2    &  GC$^o$, ALC$^o$ & GC$^o$ & 2 \\ \hline
3    & GC$^o$, $\mathcal{P}_2^o$, ALC$^o$ &GC$^o$  & 3  \\ \hline
4    &GC$^o$,  TTC$^o$, ALC$^o$ & GC$^o$ & 4 \\ \hline
$>4$ & ALC$^o$ & GC$^o$ & $n$  \\ \hline
\end{tabular}
\vspace{1mm}
\caption{NE-partitions, SO-partitions and $P_{oA}$ without Adamant Player}
\label{tab_without_adv}
\vspace{-6mm}
\end{table}
With adamant player absent, the players naturally  derive larger shares. However, we observe (from all the tables) that $P_{oA}$ is larger without adamant player. In all the cases, the $P_{oA}$ with adamant player increases to that without adamant player either when $\eta \to \infty$ or when $\eta \to 0$.

\input{Conclusions}

\label{conclusion}

\TR{
{\Large The proofs are in the next page.}
\newpage

\onecolumn}{}

\newcommand{\II}{k}
\newcommand{\sss}{{\bf s}}

\section*{Appendix A}
\label{Appendix_A}

{\bf Proof of Theorem \ref{thm:Thm1}:}
The utility of a coalition $S_m $, given by equation   \eqref{Eqn_Util_coaltiion} and \eqref{Eqn_Util_coaltiion_adamant} can be re-written as:
\begin{eqnarray*}
\varphi_{S_m} &\hspace{-2mm}=\hspace{-2mm}& \frac{  \lambda{\bar a}_{m} }{  \lambda_0a_0 
+  \lambda\sum_{l =1}^k {\bar a}_l} - \gamma {\bar a}_m , \mbox{ where aggregate actions, } \\
 {\bar a}_m &\hspace{-2mm}:=\hspace{-2mm}&  \sum_{j \in  S_m}a_j \mbox{ for each } 1\leq m \leq k \mbox{ and, } \\
 	\varphi_{S_0} &\hspace{-2mm}=\hspace{-2mm}& \frac{  \lambda_0a_0 }{  \lambda_0a_0 
 	+  \lambda\sum_{l =1}^k {\bar a}_l} - \gamma a_0.
\end{eqnarray*}
 Now, our game is reduced to a similar game as studied in \cite{dhounchak2019participate} (with action of each coalition given by the aggregate action) and the result follows from \cite[Theorem 1]{dhounchak2019participate}; the aggregate actions at NE are given by \eqref{Eqn_alstar} and the utilities are given by equations \eqref{Eqn_USm} and \eqref{NE_util_adamant}. 
 
  Any action profile, in which the aggregate actions  of each coalition equals \eqref{Eqn_alstar} forms a NE for RSG. Thus one can  have multiple NE, but the aggregate actions  and utility of each coalition  are the same at  all NE. 
 \eop

\TR{}{ }
\textbf{Proof of Theorem \ref{Thm_No_MPs_for_grt_n}:} 
Consider a strategy profile $\underbar{x}$ which leads to multiple partitions. Then as in \eqref{min_util_mult_partition} we define utility of a player to be the minimum utility among all the possible partitions emerging from $\underbar{x}$.

Let $(k_m+1)$ be the size of the biggest partition{\footnote{It can be seen from equation \eqref{Eqn_USm_player} that the utility of players decreases when the partition size, i.e., $k$ increases.}}  emerging from $\underbar{x}$ (call it ${\cal P}^*$), i.e.,  $k_m+1  =  \max_{{\cal P} ({\small{\underbar{x}}}) } |{\cal P} (\underbar{x})|$. 
$$
{\cal P}^* = \{S_0, S_1, \cdots,   S_l , S_{l+1}  \cdots\}, \mbox{ with } |{\cal P}^*|  = k+1.
$$
Now, if suppose player $i$ in coalition $S_l$ of size $m_m>1$ (we can always find such a player since otherwise all players are alone in this partition and we cannot have multiple partitions  because  of \eqref{Eqn_Coalition_rule} and \eqref{Eqn_Max_Coalition_rule})  deviates unilaterally to the strategy of being alone, i.e.,  to $\{i\}$ (changing strategy profile to $\underbar{x}'$), then we can have a partition with size at maximum  $k_m + 2$, call it $\mathcal{P}^*_{-i}$ (after splitting as in Lemma \ref{Lem_partiton_uni_dev}; since remaining players in $S_l/\{i\}$ may merge with some other coalition keeping the partition size intact).
$$
{\cal P}^*_{-i} \hspace{-0.5mm} = \hspace{-0.5mm} \{S_0, S_1, \cdots, \{i\}, S_l / \{i\}, S_{l+1}  \cdots\} \mbox{ with } |{\cal P}^*_{-i}| \hspace{-1mm}  = k+2.
$$ We have three cases based on adamant player:

\textbf{Case 1: When $\eta > 1 - 1/ (k_{m}+1)$:}  In this case the  adamant player gets non-zero utility in both the partitions, i.e.,  partition with $k_m+1$ as well as $k_m+2$ coalitions (see \eqref{NE_util_adamant}).
Then, utility of player $i$ with strategy profile  $\underbar{x}'$(using \eqref{Eqn_USm}),

{\small
\vspace{-3mm}
\begin{equation}
U_i (\underbar{x}') \ge  \frac{\lambda^2 }{ (\lambda + (k_m+1) \lambda_0)^2} = \frac{1}{(1+(k_m+1)\eta)^2}.
\label{min_x'}
\end{equation}}
 The inequality above follows because the utility of a player decreases with increasing number of coalitions.

The utility of same player $i$ under strategy profile  $\underbar{x}$ equals,
\vspace{-2mm}
$$
U_i ( \underbar{x}) \le   \frac{\lambda^2 }{ m_m  (\lambda + k_m  \lambda_0)^2} = \frac{1}{m_m(1+k_m\eta)^2},
$$
 since  the utility of a player is defined to be the minimum utility among all possible partitions and $\min(z_1,z_2,\cdots,z_n) \hspace{-1mm} \leq z_i \, \forall \, i$.
From \eqref{min_x'} and 
 Lemma \ref{Lemma_Partition_weak},  we have   (as $m_m \hspace{-1mm}> \hspace{-0.5mm} 1$ and because one can't have $m_m =2 $ and $k_m= 2$ simultaneously  for $n > 4$){\footnote{also see Corollary \ref{corollary_ALC_weak} for more details;}}:
$$
 U_i (\underbar{x}) \leq \frac{1}{m_m(1+k_m\eta)^2} <    \frac{1}{(1+(k_m+1)\eta)^2} \leq  U_i (\underbar{x}'),
 $$
 % i.e., we have, $  U_i (\underbar{x}) < U_i (\underbar{x}').$

\textbf{Case 2: When $ 1 - 1/ k_{m} \le \eta <  1 - 1/( k_{m}+1)$:}  From \eqref{NE_util_adamant},  the adamant player gets non-zero utility in partition with $k_m+1$ coalitions but zero utility with $k_m+2$ coalitions.
%Then, we have
%$$
%\frac{k_m-1}{k_m} \leq \eta \leq \frac{k_m}{k_m+1}
%$$
Once again, from \eqref{Eqn_USm} the
utility of player $i$ with strategy profile $\underbar{x}'$ (adversary insignificant),
\vspace{-2mm}
\begin{equation}
\label{adv_zero_one}
U_i(\underbar{x}')  \ge   \Big( \frac{1}{k_m+1}\Big)^2.
\vspace{-1mm}
\end{equation}
Thus the utility of player $i$ with strategy profile  $\underbar{x}$ equals (inequality as explained in Case 1), 
\vspace{-1mm}
\begin{equation*}
U_i(\underbar{x})  \leq \frac{1}{m_m(1+ k_m\eta)^2}.
\vspace{-1mm}
\end{equation*}
By the conditions of Case 2, we have
  $k_m\eta \geq (k_m-1)$ and  as in Lemma \ref{Lemma_Partition_weak} we have  $ \sqrt{m_m} k_m  >  k_m+1$ (as explained in Case 1):
  \vspace{-1mm}
\begin{equation*}
\sqrt{m_m}(1+ k_m\eta) \ge  \sqrt{m_m} k_m  >  k_m+1. 
\vspace{-2mm}
\end{equation*}  
Hence 
from \eqref{adv_zero_one}: 
\vspace{-1mm}
\begin{equation*}
U_i(\underbar{x})   \leq  \frac{1}{m_m(1+ k_m\eta)^2} \ < \  \Big( \frac{1}{k_m+1}\Big)^2  \le U_i(\underbar{x}') .
%\\
%(k_m+1) & < & \sqrt{m_m}( k_m\eta+1)
\end{equation*}

%\vspace{-1mm}
\textbf{Case 3: When adamant player gets zero utility in both the partitions:}
Once again, the
utility of player $i$ with strategy profile $\underbar{x}'$,
\vspace{-1mm}
\begin{equation}
\label{adv_zero_two}
U_i(\underbar{x}')  \ge   \Big( \frac{1}{k_m+1}\Big)^2.
\vspace{-1mm}
\end{equation}
%
% emerging from $\underbar{x}$, then this partition also emerges from $\underbar{x}'$ and hence the minimum with  $\underbar{x}$  is either less than or equal to that with  $\underbar{x}'$, i.e., . $U_i (\underbar{x}') \ge   U_i (\underbar{x}) $.
  As before:
\begin{equation*}
%\vspace{-1mm}
U_i(\underbar{x})  \leq \frac{1}{m_mk_m^2}. %\vspace{-1mm}
\end{equation*}
As in Lemma \ref{Lemma_Partition_weak}, $\sqrt{m_m} k_m  >  (k_m + 1)$ and hence  
%\vspace{-1mm}
\begin{equation*}
U_i(\underbar{x})   \leq  \frac{1}{m_mk_m^2}   <   \Big( \frac{1}{k_m+1}\Big)^2  \le  U_i(\underbar{x}') .
\end{equation*}
 Thus, player $i$  player finds it strictly better to deviate and hence the result. 
\eop

\medskip

\textbf{ Proof of Corollary \ref{corollary_ALC_weak}:} Consider any partition ${\cal  P}$ other than ALC. Let $m^*$ be the size of the biggest coalition  of ${\cal  P}$.  Then $m^* \ge 2$.  If $m^* = 2$, then $k := |{\cal P}| - 1 \ge  \lceil n/2 \rceil $ (lower bound achieved  when maximum coalitions are exactly of size  2). Thus   $m^* k^2  > (k+1)^2$, as $n > 4$  (note as $k$ increases, $(k+1)^2/k^2$ decreases).  If  $2 <  m^* < n/2$, then  $k := |{\cal P}| - 1 \ge  \lceil n/m^* \rceil$ and hence:

\vspace{-4mm}
{\small \begin{eqnarray*}
\left ( \frac{k+1}{k}   \right )^2  =  \left (  1 + \frac{1}{k}  \right )^2 & \le &  \left (  1 +  \frac{m^*}{n} \right )^2 \le  (1.5)^2  <  3 \le m^*.
\end{eqnarray*}}
For $m^* > n/2$, we have $k \ge 2${\footnote{$k=1$ refers to the grand coalition of players which is never possible for $n>4$ (check Lemma \ref{Lemma_Partition_weak} conditions)} and hence 
$$
 \Big(1+\frac{1}{k}\Big)^2 \le  \Big(1+\frac{1}{2}\Big)^2 = 2.25  <  \frac{n}{2}  < m^*.
$$
since $n>4$. Thus conditions of Lemma   \ref{Lemma_Partition_weak} are satisfied for all partitions other than ALC/ALC$^o$ and hence the result. \eop

\section*{Appendix B}
\label{Appendix}

\textbf{Proof of Lemma \ref{Lem_partiton_uni_dev}:} We prove it in two steps: i) $\underbar{x}' \to {\cal P}_{-i}$ and  ii) $\underbar{x}' \to !\mathcal{P}_{-i}$. 

To prove $\underbar{x}' \hspace{-1mm} \to \hspace{-1mm} \mathcal{P}_{-i}$:  it is clear by definition that  every coalition of $ \mathcal{P}_{-i}$ satisfies the    requirement \eqref{Eqn_Coalition_rule}  (with  $\underbar{x}'$). Hence, it suffices to prove that it is minimal  as in \eqref{Eqn_Max_Coalition_rule}.
 
If possible consider a (better)  partition $\mathcal{P}'$ which satisfies \eqref{Eqn_Coalition_rule} and  such that $ \mathcal{P}'  \prec \mathcal{P}_{-i}$.  This means, from \eqref{Eqn_Max_Coalition_rule},   there exist at least a pair of coalitions $S_1, S_2  \in  \mathcal{P}_{-i}$ and an $S \in \mathcal{P}'$ such that $S_1 \cup S_2 \subset S$. 
\Cmnt{One may  have more such pairs of $S_1$, $S_2$ (and $S$) or some times more than two such coalitions  merging  together to a common coalition of $ \mathcal{P}'$.} Observe that $\{i\} \in \mathcal{P}'
\cap \mathcal{P}_{-i}$, 
as player $i$ deviates unilaterally to $\{i\}.$

If  all such merging coalitions in $\mathcal{P}_{-i}$ are not equal to  $S_{l}\backslash \{i\}$ the merging coalitions will  also belong  to $\mathcal{P}$  (i.e., for example  if 
  $S_1 \neq S_2 \neq S_{l}\backslash \{i\}$, then $S_1, S_2 $ also belong  to $\mathcal{P}$),    then  one can  construct a  better partition\footnote{Partition ${\cal P}''$ 
contains all coalitions of  ${\cal P}'$, except that $\{i \}$  and $S_{l}\backslash \{i\}$ are merged in  ${\cal P}''$.
 } ${\cal P}'' \prec {\cal P}$ and   $\underbar{x} \to {\cal P}''$,   which contradicts  $\underbar{x}  \to \mathcal{P}$.  

 On the other hand, if one of the merging coalitions equal $S_{l}\backslash{\{i\}}$,  then  ${\cal P}'$ is not comparable  with  ${\cal P}$ as in \eqref{Eqn_Max_Coalition_rule_0}  (i.e., neither is better than the other), as  $\hspace{-.5mm} \{i\} \hspace{-1mm} \in \hspace{-1mm} {\cal P}'$. 
Further  ${\cal P}' \hspace{-1mm}$ satisfies  \eqref{Eqn_Coalition_rule} with  $\underbar{x}$ and hence
 $\underbar{x} \hspace{-1mm}\to \hspace{-1mm} {\cal P}'$. That means $\underbar{x}$   leads to multiple partitions and this contradicts the hypothesis that    $\underbar{x} \hspace{-1mm}\to  !{\cal P}$.
This proves (i). 
 
 Next we prove uniqueness in (ii). % which says that the new resulting profile also leads to a unique partition $\mathcal{P}_{-i}$.
If possible  
  $\underbar{x}'$ leads to multiple partitions, say $\mathcal{P}_{-i}$ (defined in hypothesis) and   $\mathcal{P}'$. 
This implies   $\mathcal{P}_{-i}$ is not comparable to  $\mathcal{P}'$. Further  observe  $ \{i\} \in \mathcal{P}'$ and hence  $ \mathcal{P}'$ is not even comparable to $  \mathcal{P}$. 
Further more,   it is easy to verify that any coalition  that satisfies \eqref{Eqn_Coalition_rule} with   $\underbar{x}'$ also satisfies    \eqref{Eqn_Coalition_rule} with   $\underbar{x}$.  In all we have that 
 $\underbar{x} \to {\cal P}'$,   which again contradicts the uniqueness of  $\underbar{x}  \to ! \mathcal{P}$.   \eop

\textbf{Proof of Lemma \ref{Lemma_Partition_weak}:}
Wlog we can assume that the $n$ C-players (i.e., with influence factor $\lambda$) form $k$ coalitions where $k\leq n$ , i.e., 
\TR{
\begin{eqnarray*}
\mathcal{P} &=& \{\{0\},\{1,\cdots,m_1\},\{m_1+1,\cdots,m_2\},\cdots,\{m_{k-1}+1,\cdots,m_k\}\}.
\end{eqnarray*}}
{\vspace{-2mm} 
	\begin{eqnarray*}
\mathcal{P} &\hspace{-2mm}= \hspace{-2mm}& \{\{0\},\{1,\cdots,m_1\},\{m_1+1,\cdots,m_2\},\cdots,\\
 &&\{m_{k-1}+1,\cdots,m_k\}\}.
 \vspace{-2mm}
\end{eqnarray*}}
\TR{
Hence, we have $m_1,\,(m_2-m_1),\cdots,(m_k-m_{k-1})$ number of players in each of the $k$ coalitions of C-players respectively where $m_k = n$.}{}

Consider the best response of (say $m_1 \hspace{-1mm}=\hspace{-1mm} m^*$) player $1$ against any strategy profile $\underbar{x} \hspace{-1mm} \to !\mathcal{P}$; player $1$ could either choose to remain alone (i.e., $x_1 \hspace{-1mm}= \hspace{-1mm} \{1\}$) or could form coalition with all or a subset of $(m_1-1)$ players (i.e., $x_1 \subset  \{1,\cdots,m_1\}$) resulting into a new strategy profile $\underbar{x}'$. In particular, we would show that forming coalition with all  players (as given by $\underbar{x} \to !\mathcal{P}$)
is strictly inferior to remaining alone, i.e., player $1$ could get higher utility by unilaterally deviating to $\{1\}$.

\textbf{Case 1: When} $\eta > 1-1/(k+1)$: In this case the adamant player gets non-zero utility in both the partitions, i.e., partition with $k+1$ as well as $k+2$ coalitions.

 Then, from \eqref{Eqn_USm} utility of player 1 when it chooses to remain alone (with strategies of the others remaining the same),
 %\vspace{-1mm}
\begin{equation}
    U_1(\underbar{x}') = \Big(\frac{\lambda}{\lambda+(k+1)\lambda_0}\Big)^2  = \Big(\frac{1}{1+(k+1)\eta}\Big)^2.
\label{eq:Eq4.21}
\end{equation}
Similarly, utility of  player 1 when it proposes to form coalition with all $(m_1-1)$ players,
%\vspace{-1.5mm}
\begin{equation*}
    U_{1}(\underbar{x})=   \frac{1}{m_1}\Big(\frac{\lambda}{\lambda+k\lambda_0}\Big)^2 =  \frac{1}{m_1}\Big(\frac{1}{1+k\eta}\Big)^2.
\label{eq:Eq4.22}
\end{equation*}

\Cmnt{We have $m_1 >1$ and as in Lemma \ref{Lemma_Partition_weak}, we have  $ \sqrt{m_1}k  >  k+1$:\vspace{-3mm}
\begin{eqnarray*}
\sqrt{m_1}(1+k\eta) \ge  1+(k+1)\eta.
\end{eqnarray*} } 
Since $m_1 >1$, 
from \eqref{eq:Eq4.21}, player $i$ finds it better to deviate: 
\vspace{-7mm}

{\small \begin{equation*}
U_1(\underbar{x}')   =   \Big(\frac{1}{1+(k+1)\eta}\Big)^2 \ > \  \frac{1}{m_1}\Big(\frac{1}{1+k\eta}\Big)^2 = U_1(\underbar{x}) .
%\\
%(k_m+1) & < & \sqrt{m_m}( k_m\eta+1)
%\vspace{-1mm}
\end{equation*}}
if $ \sqrt{m_1}k  >  k+1$.

 %Since, this is valid for all coalitions hence it is sufficient to check for $m^* := \max_{S_i \in \mathcal{P}} |S_i|$
%\vspace{-1mm}
\textbf{Case 2: When $ 1 - 1/ k \le \eta <  1 - 1/( k+1)$:}  From \eqref{NE_util_adamant},  the adamant player gets non-zero utility in partition with $k+1$ coalitions but zero utility with $k+2$ coalitions.

Now, utility of player 1 when it proposes to form coalitions with $(m_1-1)$ players,
\vspace{-2mm}
\begin{eqnarray}
\label{case2_coa}
U_1(\underbar{x})& = & \frac{1}{m_1}\Big( \frac{1}{1+k\eta}\Big)^2.
\end{eqnarray}
When player 1 chooses to remain alone (with strategies of other players remaining the same) then, utility of player 1 is given by
\vspace{-5mm}
\begin{eqnarray}
U_1(\underbar{x}') & = & \Big( \frac{1}{k+1}\Big)^2.
\label{case2_alone}
\end{eqnarray}

By the conditions of Case 2 we have $k\eta \geq (k-1)$  %and as in Lemma \ref{Lemma_Partition_weak}, we have  $ \sqrt{m_1}k  >  k+1$:
\begin{eqnarray*}
\sqrt{m_1}(1+k\eta) \ge  \sqrt{m_1}(1+k-1) =\sqrt{m_1}k .
\end{eqnarray*}  
Hence 
from \eqref{case2_alone}: 
\vspace{-1mm}
{ \small \begin{eqnarray*}
U_1(\underbar{x}')   = \Big( \frac{1}{k+1}\Big)^2 \ > \   \frac{1}{m_1}\Big( \frac{1}{k}\Big)^2 \geq \   \frac{1}{m_1}\Big( \frac{1}{1+k\eta}\Big)^2 = U_1(\underbar{x}),
%\\
%(k_m+1) & < & \sqrt{m_m}( k_m\eta+1)
\end{eqnarray*}}
if $ \sqrt{m_1}k  >  k+1$.

\textbf{Case 3: When adamant player gets zero utility in both the partitions}

Once again the utility of player 1 when it proposes to form coalitions with $(m_1-1)$ players,
\vspace{-2mm}
\begin{eqnarray}
\label{case3_coa}
U_1(\underbar{x}) & = & \frac{1}{m_1}\Big( \frac{1}{k}\Big)^2.
\end{eqnarray}
When player 1 chooses to remain alone (with strategies of other players remaining the same) then, utility of player 1 is given by
\vspace{-4mm}
\begin{eqnarray}
\label{case3_alone}
U_1(\underbar{x}') & = & \Big( \frac{1}{k+1}\Big)^2.
\end{eqnarray}

Hence 
from \eqref{case3_alone}: 
\begin{eqnarray*}
\hspace{12mm} U_1(\underbar{x}')   = \Big( \frac{1}{k+1}\Big)^2 \ > \ \frac{1}{m_1}\Big( \frac{1}{k}\Big)^2  = \ U_1(\underbar{x}) . 
%\\
%(k_m+1) & < & \sqrt{m_m}( k_m\eta+1) \vspace{-2mm}
%\hspace{16mm}\mbox{\eop}
\end{eqnarray*}
if $ \sqrt{m_1}k  >  k+1$. \eop
 
\TR{
\textbf{Proof of Lemma \ref{Lemma_weak_NE}:} 
Since the given partition $\mathcal{P}$ is weak,  by definition  for any strategy profile $\underbar{x} \to !\mathcal{P}$,    there exists  a player $i$   which  gets higher utility at its $i$-$u.d.p.$  Hence, any strategy profile that leads to partition $\mathcal{P}$ uniquely,   cannot be a NE. 
Also, since multiple partitions are not possible at NE  (given) hence $\mathcal{P}$ cannot   result from a strategy profile which leads to multiple partitions and simultaneously be  partition at NE. Hence,  $\mathcal{P}$ can never emerge from a NE. \eop }{}

 {
{\bf Proof of Lemma \ref{Lemma_SO_Partition}:}
Let $\eta \ge 1$ (adamant player gets non-zero utility in all such partitions). 
From \eqref{Eqn_USm}, one can verify  ($\{m_i\}$-sizes of coalition, $k+1$-size of partition):

\vspace{-5mm}
{\small
\begin{eqnarray}
U_{SO}^*:= \max_{\mathcal{P}} \sum_{S_i \in \mathcal{P};i \neq 0} U_{S_i}  \hspace{-2mm} & \hspace{-2mm}  = \hspace{-2mm} &  \hspace{-2mm} 
    {  \max_{ \{ \{ m_i \}_{i \le k} , k  \} }  \sum_{i=1}^k \sum_{j=1}^{m_i}  \frac{ \lambda^2 }{m_i (\lambda +  k \lambda_0)^2 }  }.  \nonumber \\
    & &  \hspace{-35mm} =\     {  \max_{ \{ \{ m_i \}_{i \le k} , k  \} }  \sum_{i=1}^k \frac{ \lambda^2 }{ (\lambda +  k \lambda_0)^2 }  }  \ = \ 
 {   \max_{1\le   k   \le n}   \frac{ k  \lambda^2 }{ (\lambda +  k \lambda_0)^2 }  }.  \label{Eqn_USO_initial}
\end{eqnarray}}
One can equivalently  minimize:
\begin{eqnarray*}
 \min_{1\le   k   \le n}   \frac{ (\lambda +  k \lambda_0)^2 }{ k  \lambda^2 } &=& \frac{1}{\lambda^2} \min_{1\le   k   \le n} \frac{\lambda^2+2k\lambda_0\lambda+k^2\lambda_0^2}{k}. 
 %\\
% & = &  \frac{1}{\lambda^2} \min_{1\le   k   \le n} \frac{\lambda^2   } {k } + 2\lambda_0\lambda + k  \lambda_0^2 
\end{eqnarray*}or equivalently consider:
\vspace{-2mm}
\begin{eqnarray}
\label{Eqn_Convex_function}
\min_{1\le   k   \le n}   \frac {  \lambda^2   } {k }   + k  \lambda_0^2 .
\end{eqnarray}
By relaxing $k$ to real numbers, and equating the derivative to zero  (verify the second derivative is positive) we obtain:\vspace{-2mm}
\begin{eqnarray*}
 -\frac{\lambda^2}{k^2} + \lambda_0^2   = 0   \mbox{ or }  
 k^*   =    1/\eta. 
\end{eqnarray*} 
This implies (by convexity) that the optimizer among integers  is  $k^* = 1$ when $\eta \ge 1$, i.e., GC is the SO-partition. 
On the other hand, when $\eta \le  1- 1/2 = 0.5$,  from \eqref{NE_util_adamant} the adversary gets insignificant in all partitions other than GC, and one needs to maximize 
\vspace{-2mm}
\TR{ \begin{eqnarray*}
U_{SO}^* &=& \max \left \{  \frac{\lambda^2}{(\lambda+\lambda_0)^2} ,      \max_{ \{ \{ m_i \}_{i \le k} , k >  1 \} }  \sum_{i =1}^k  \frac{ 1 }{  k^2 }   \right \}.  %\TR { }{}
= \max \left \{  \frac{\lambda^2}{(\lambda+\lambda_0)^2} ,      \max_{ \{ \{ m_i \}_{i \le k} , k >  1 \} }   \frac{ 1 }{  k }   \right \}.  =
 \max \left \{  \frac{\lambda^2}{(\lambda+\lambda_0)^2} ,     \frac{1}{2}   \right \} .
\end{eqnarray*} }{\begin{eqnarray*}
U_{SO}^* &=& \max \left \{  \frac{\lambda^2}{(\lambda+\lambda_0)^2} ,      \max_{ \{ \{ m_i \}_{i \le k} , k >  1 \} }  \sum_{i =1}^k  \frac{ 1 }{  k^2 }   \right \}.  \\
&=& \max \left \{  \frac{\lambda^2}{(\lambda+\lambda_0)^2} ,      \max_{ \{ \{ m_i \}_{i \le k} , k >  1 \} }   \frac{ 1 }{  k }   \right \}. \\  &=&
\max \left \{  \frac{\lambda^2}{(\lambda+\lambda_0)^2} ,     \frac{1}{2}   \right \} .
\end{eqnarray*}}
When $\eta \leq   \sqrt{2}-1= 0.414 $,  GC is the  SO-partition  and for $ 0.414   \le \eta \leq 0.5$ any  ${\cal P}^*_2$ is an SO-partition, which completes the proof of part (ii).  
%\TR{\\
Similarly when $\eta  \le  1- 1/3$ we have
\vspace{-1mm}
\begin{eqnarray*}
{\bar U}^*_{SO}
= \max \left \{  \frac{\lambda^2}{(\lambda+\lambda_0)^2} ,   \frac{2\lambda^2}{(\lambda+2\lambda_0)^2} ,      \frac{1}{3}  \right \} .
\end{eqnarray*} 
Progressing this way, for any $\eta \le 1 - 1/k$  (as in \eqref{Eqn_USO_initial}): 
\begin{eqnarray*}
{\bar U}^*_{SO}
= %\left \{ 
\hspace{-2mm}  \begin{array}{llll}
 \max \left \{ \max_{k' < k } \frac{ k' \lambda^2}{(\lambda+ k' \lambda_0)^2}  , \   \frac{1}{k}  \right \}   & \mbox{for any }  k \le  n,   \\
%
% \max \left \{ \max_{k' < k } \frac{ k' \lambda^2}{(\lambda+ k' \lambda_0)^2}    \right \}  & \mbox{for   }  k >  n .
\end{array} %\right . 
\end{eqnarray*}
%
% (by convexity) optimizer among integers:
%
% 
%\begin{eqnarray*}
% k^*=\min\left \{ n,   \max \left  \{1 ,   \left \lfloor 1/ \eta \right   \rfloor  \right  \}\right  \}  \text{ or }  \\
%\min\left \{ n,  \max \left  \{1 ,   \left \lceil  1/\eta \right   \rceil  \right  \}\right  \}   \text{ or }     \left   \lceil \frac {1}{1-\eta}  \right \rceil 
%\end{eqnarray*}
%{\color{red}
%The last term should be
%$$
%{k} = \lceil \frac{1}{1-\eta} \rceil
%$$
%}
But with $\eta  > 1/2$, the relaxed $k^* = 1/ \eta  < 2$.  Thus  by convexity of \eqref{Eqn_Convex_function} the maximizer of the first term among integers is either at 1 or 2, i.e., when $\eta \le 1 - 1/k$

\vspace{-4mm}
%{\small
\begin{eqnarray*}
{\bar U}^*_{SO}
=%\left \{ 
\begin{array}{lll}
\max \left \{  \frac{\lambda^2}{(\lambda+\lambda_0)^2} ,   \frac{2\lambda^2}{(\lambda+2\lambda_0)^2} ,   \frac{1} {  k }  \right \} & \mbox{ \normalsize if }  
 k \le  n, \\
%\max \left \{  \frac{\lambda^2}{(\lambda+\lambda_0)^2} ,   \frac{2\lambda^2}{(\lambda+2\lambda_0)^2}   \right \}  & \mbox{\normalsize else. } 
\end{array}  %\right .
\end{eqnarray*}
We have GC is best among the first two if 
\vspace{-2mm}
$$
1+ 2\eta  \ge    \sqrt{2} (1+\eta)  \mbox{ or if  } \eta \ge   1/ \sqrt{2}  = 0.707.
\vspace{-2mm}
$$further in this range for all $n$,  GC is better than the third possibility also (if it is feasible). In a similar way one can prove that ${\cal P}_2$ is optimal for all other values of $\eta$.
Thus we proved the Lemma. \eop}{}

{\bf Proof of Lemma \ref{Lemma_SO_no_adv}:}
Consider $\{m_i\}$ to be the size of coalition $S_i$, i.e.,$|S_i|$ and $k$ be the size of partition $\mathcal{P}^o$, i.e. , $|\mathcal{P}^o|$.
\vspace{-5mm}
\begin{eqnarray}
\hspace{3mm} U_{SO}^*:= \max_{\mathcal{P}^o} \sum_{S_i \in \mathcal{P}^o} U_{S_i}  \hspace{-2mm} & \hspace{-2mm}  = \hspace{-2mm} &  \hspace{-2mm} 
    {  \max_{ \{ \{ m_i \}_{i \le k} , k  \} }  \sum_{i=1}^k \sum_{j=1}^{m_i} \frac{1}{m_ik^2}  }.  \nonumber \\
    & &  \hspace{-30mm} =\     {  \max_{ \{ \{ m_i \}_{i \le k} , k  \} }  \sum_{i=1}^k \frac{1}{k^2}  }  \ = \ 
 {   \max_{1\le   k   \le n}   \frac{1}{k}  }.  \label{Eqn_USO_initial_no_adv}
\end{eqnarray}
Since the minimum possible value of $k$ is 1, we have GC$^o$ is the only SO-partition in this case. \eop

\TR{\section*{Computations for $n\le4$ with adversary}
\label{comp_small_with_adv}

\subsection*{ When $n=2$:}
To compute NE, we consider \emph{best response} (BR) of player 1 against strategy $\{1,2\}$ of player 2.

From equation \eqref{Eqn_USm_player}, utility of player 1 when it chooses $\{1,2\}$, we have
\begin{eqnarray}
\label{util_gc_2}
 \frac{1}{2}\Big(\frac{\lambda}{\lambda+\lambda_0}\Big)^2 =   \frac{1}{2}\Big(\frac{1}{1+\eta}\Big)^2.
\end{eqnarray}
Similarly, utility of player 1 when it chooses $\{1\}$, we have
\begin{eqnarray}
\label{util_alc_2}
\Big(\frac{\lambda}{\lambda+2\lambda_0}\Big)^2 = \Big(\frac{1}{1+2\eta}\Big)^2.
\end{eqnarray}
Comparing equation \eqref{util_gc_2} and \eqref{util_alc_2}, GC is formed ,i.e., $\{1,2\}$ lies in best response of player 1 against $\{1,2\}$ strategy of player 2, if
\begin{eqnarray*}
\frac{1}{2}\Big(\frac{1}{1+\eta}\Big)^2 & \geq &  \Big(\frac{1}{1+2\eta}\Big)^2 \mbox{ , i.e.,  if, }  \\
({1+2\eta}) & \geq & \sqrt{2}({1+\eta})  \mbox{ , i.e.,  if, }   \\
(2-\sqrt{2})\eta & \geq & (\sqrt{2}-1) \mbox{ , i.e.,  if, }  \\
\eta & \geq & 0.707.
\end{eqnarray*} 
Hence for the range $\eta \geq0.707 $, GC gives maximum utility to the players. Further, 
one can check that adversary is significant in both possible partitions; thus $GC$ is a NE-partition.  Recall ALC/ALC$^o$ is always a NE.

When $0.5 < \eta <0.707$, adversary remains significant and hence ALC is the only NE in this range. When $\eta \leq 0.5$, adversary becomes insignificant when ALC is formed and hence, we need to compare utilities of players at GC and ALC$^o$.
Utility of player 1 when he chooses $\{1\}$ against $\{1,2\}$ strategy of player 2,
\begin{eqnarray}
\label{alc_insig_adv_2}
 \frac{1}{2^2} = \frac{1}{4}.
\end{eqnarray}
Comparing equation \eqref{util_gc_2} and \eqref{alc_insig_adv_2}, ALC$^o$ is formed ,i.e., $\{1\}$ lies in best response of player 1 against $\{1,2\}$ strategy of player 2, if
\begin{eqnarray*}
\frac{1}{4}  & \geq & \frac{1}{2}\Big(\frac{1}{1+\eta}\Big)^2 \mbox{ , i.e.,  if, }   \\
(1+\eta) & \geq &  \sqrt{2}\mbox{ , i.e.,  if, } \\
\eta & \geq & 0.414.
\end{eqnarray*}
Hence, for the range $0.414 \leq \eta  \leq  0.5$ and $ 0 \leq \eta < 0.414$, ALC$^o$ and GC are formed respectively. The SO-partitions can be calculated from Lemma \ref{Lemma_SO_Partition}.
\subsection*{ When $n=3$:}
We again compute BRs directly. Consider the case when player 2  and 3 chooses $GC = \{1,2,3\}$. Player 1 could choose strategy GC, $\{1,2\}$ (or equivalently $\{1,3\}$) or $\{1\}$. From
 \eqref{Eqn_USm_player} the utility of player 1  if  it
 chooses  $GC = \{1,2, 3\}$ (partition GC is formed) 
is 
\begin{eqnarray}
\label{util_gc_3}
 \frac{1}{3}\Big(\frac{\lambda}{\lambda+\lambda_0}\Big)^2 =   \frac{1}{3}\Big(\frac{1}{1+\eta}\Big)^2.
\end{eqnarray}
From equation \eqref{min_util_mult_partition}, utility of player 1  if  it
 chooses  $ \{1,2\}$ (multiple partitions are formed) 
is:
 \begin{eqnarray}
\label{util_mult_3}
 \frac{1}{2}\Big(\frac{\lambda}{\lambda+2\lambda_0}\Big)^2 =   \frac{1}{2}\Big(\frac{1}{1+2\eta}\Big)^2.
\end{eqnarray}
Next, utility of player 1  if  it
 chooses  $ \{1\}$ ($\mathcal{P}_2$ type partition is formed) 
is:
 \begin{eqnarray}
\label{util_alone_3}
\Big(\frac{\lambda}{\lambda+2\lambda_0}\Big)^2 =   \Big(\frac{1}{1+2\eta}\Big)^2.
\end{eqnarray}
From equations \eqref{util_mult_3} and \eqref{util_alone_3}, player 1 gets strictly better utilities when it chooses $\{1\}$. Thus, comparing equation \eqref{util_gc_3} and \eqref{util_alone_3}, $\{1\}$ lies in best response of player 1 against $\{1,2,3\}$ strategy of player 2 and 3, if
\begin{eqnarray*}
 \Big(\frac{1}{1+2\eta}\Big)^2 & \geq &   \frac{1}{3}\Big(\frac{1}{1+\eta}\Big)^2 \mbox{ , i.e.,  if, } \\
 \sqrt{3}(1+\eta) & \geq & (1+2\eta) \mbox{ , i.e.,  if, } \\
 (\sqrt{3}-1) & \geq & (2-\sqrt{3})\eta \mbox{ , i.e.,  if, }  \\
 2.732 & \geq & \eta.
\end{eqnarray*}
 Thus, GC is a NE-partition when  $\eta > 2.732$. One can check that adversary is significant in all possible partitions in this range. ${\cal P}_2$ can also be a NE-partition in this range for some strategy profiles. For example, $x_1 = \{1\}$, $x_2= \{2,3\}$ and $x_3= \{2,3\}$.

When player 1 and 3 chooses $\{1\}$ and $\{1,2,3\}$ respectively, player 2 could either form coalition with player 3 or remain alone. From \eqref{Eqn_USm_player} the utility of player 2  if  it chooses  $\{1,2,3\}$ (or equivalently $\{2,3\}$) is given by equation \eqref{util_mult_3}.
Similarly, the utility of player 2  if  it chooses  $\{2\}$ (or equivalently $\{1,2\}$) is
\begin{eqnarray}
\label{util_alc_3}
\Big(\frac{\lambda}{\lambda+3\lambda_0}\Big)^2 =   \Big(\frac{1}{1+3\eta}\Big)^2.
\end{eqnarray}
Comparing equation \eqref{util_mult_3} and \eqref{util_alc_3}, ALC is formed ,i.e., $\{2\}$ lies in best response of player 2 against $\{1\}$ and $\{1,2,3\}$ strategy of player 1 and 3, if
\begin{eqnarray*}
   \Big(\frac{1}{1+3\eta}\Big)^2 & \geq & \frac{1}{2}\Big(\frac{1}{1+2\eta}\Big)^2 \mbox{ , i.e.,  if, } \\
   \sqrt{2}(1+2\eta) & \geq & (1+3\eta) \mbox{ , i.e.,  if, } \\
   (\sqrt{2}-1)  & \geq & (3-2\sqrt{2})\eta \mbox{ , i.e.,  if, } \\
   2.414  & \geq & \eta.
\end{eqnarray*}
Thus, $\mathcal{P}_2$ type partitions are are NE-partitions when $2.414 < \eta  \leq 2.732$. Adversary is significant for all possible paartitions in this range as well. Also, ALC is a NE-partition until the adversary becomes insignificant , i.e., $0.57 < \eta  \leq 2.414$ (from equation \eqref{Eqn_USm}).

Similarly, when $0.5 < \eta  \leq0.57$ adversary becomes insignificant in ALC partition. Hence, we need to check player 2's best response. As explained above, we have $\mathcal{P}_2$ and ALC$^o$ are the NE-partitions in this range.

When $\eta \leq 0.5$, we have GC as the only partition where adversary is significant.
Now consider best response of player 1 against strategy $GC = \{1,2,3\}$ of player 2 and 3.

Utility of player 1 when he chooses strategy $GC = \{1,2,3\}$ is given by equation \eqref{util_gc_3}. From equation \eqref{min_util_mult_partition} utility of player 1 if it chooses $\{1,2\}$ (or equivalently $\{1,3\}$ is:
\begin{eqnarray}
\frac{1}{2}\frac{1}{2^2} &=& \frac{1}{8}.
\label{util_p2_3}
\end{eqnarray}
Similarly, utility of player 1 if it chooses $\{1\}$ (or equivalently $\{1,3\}$ is:
\begin{eqnarray}
\label{util_p3_3}
\frac{1}{2^2} &=& \frac{1}{4}.
\label{util_p2_3}
\end{eqnarray}
From equations \eqref{util_p2_3} and \eqref{util_p3_3}, player 1 gets strictly better utilities when he chooses $\{1\}$. Thus, comparing equation \eqref{util_gc_3} and \eqref{util_p3_3}, GC is formed ,i.e., $\{1,2,3\}$ lies in best response of player 1 against $\{1,2,3\}$ strategy of player 2 and 3, if
\begin{eqnarray*}
 \frac{1}{3}\Big(\frac{1}{1+\eta}\Big)^2  & \geq &  \frac{1}{4}\mbox{ , i.e.,  if, }  \\
 2 &\geq & \sqrt{3}(1+\eta)\mbox{ , i.e.,  if, }  \\
 (2-\sqrt{3}) & \geq & \sqrt{3}\eta \mbox{ , i.e.,  if, } \\
 0.15 &\geq & \eta.
\end{eqnarray*}
Hence, GC and $\mathcal{P}_2^o$  is the NE-partition when $\eta \le 0.15$ and $0.15 < \eta \leq 0.5$ respectively. The SO-partitions can be calculated from Lemma \ref{Lemma_SO_Partition} as before.

\subsection*{ When $n=4$:}
\textbf{Case 1: When $\eta >0.75$ } In this case, all the possible partitions have significant adversary.
Here also we compute BRs directly. Consider the case when player 2,3 and 4 chooses $GC = \{1,2,3,4\}$. Player 1 could choose strategy GC, $\{1,2,3,4\}$, $\{1,2,3\}$ (or equivalently $\{1,3,4\}$ or $\{1,2,4\}$), $\{1,2\}$  (or equivalently $\{1,3\}$ or $\{1,4\}$) or $\{1\}$. From
 \eqref{Eqn_USm_player} the utility of player 1  if  it
 chooses  $GC = \{1,2, 3,4\}$ (partition GC is formed) 
is 
\begin{eqnarray}
\label{util_gc_4_adv}
 \frac{1}{4}\Big(\frac{\lambda}{\lambda+\lambda_0}\Big)^2 =   \frac{1}{4}\Big(\frac{1}{1+\eta}\Big)^2.
\end{eqnarray}
From equation \eqref{min_util_mult_partition}, utility of player 1  if  it
 chooses   $\{1,2,3\}$ (or equivalently $\{1,3,4\}$ or $\{1,2,4\}$) is 
\begin{eqnarray}
\label{util_par3_4_adv}
 \frac{1}{3}\Big(\frac{\lambda}{\lambda+2\lambda_0}\Big)^2 =   \frac{1}{3}\Big(\frac{1}{1+2\eta}\Big)^2.
\end{eqnarray}
Similarly,  utility of player 1  if  it
 chooses $\{1,2\}$  (or equivalently $\{1,3\}$ or $\{1,4\}$) is
\begin{eqnarray}
\label{util_par2_4_adv}
 \frac{1}{2}\Big(\frac{\lambda}{\lambda+2\lambda_0}\Big)^2 =   \frac{1}{2}\Big(\frac{1}{1+2\eta}\Big)^2.
\end{eqnarray}
  Utility of player 1  if  it
 chooses $\{1\}$  is
\begin{eqnarray}
\label{util_alone_4_adv}
\Big(\frac{\lambda}{\lambda+2\lambda_0}\Big)^2 =   \Big(\frac{1}{1+2\eta}\Big)^2.
\end{eqnarray}
From equations \eqref{util_par3_4_adv}, \eqref{util_par2_4_adv} and \eqref{util_alone_4_adv}, player 1 gets strictly better utilities when he chooses $\{1\}$. Thus, comparing equation \eqref{util_gc_4_adv} and \eqref{util_alone_4_adv}, GC is formed ,i.e., $\{1,2,3,4\}$ lies in best response of player 1 against $\{1,2,3,4\}$ strategy of player 2,3 and 4, if
\begin{eqnarray*}
 \frac{1}{4}\Big(\frac{1}{1+\eta}\Big)^2 & \geq & \Big(\frac{1}{1+2\eta}\Big)^2 \mbox{ , i.e.,  if, } \\
 (1+2\eta) & \geq & 2(1+\eta). 
\end{eqnarray*}
which is not possible. Hence, GC cannot be a NE-partition.

Next we consider BR of player 2 against $\{1\}$, $\{1,2,3,4\}$ and $\{1,2,3,4\}$ strategy of player 1,3 and 4 respectively.

Utility of player 2 when it chooses $\{1,2,3,4\}$ (or equivalently $\{2,3,4\}$), is
\begin{eqnarray}
\label{player_4_p2}
\frac{1}{3}\Big(\frac{\lambda}{\lambda+2\lambda_0}\Big)^2 =   \frac{1}{3}\Big(\frac{1}{1+2\eta}\Big)^2.
\end{eqnarray}

From equation \eqref{min_util_mult_partition}, utility of player 2  if  it
 chooses   $\{2,3\}$ (or equivalently $\{1,2,3\}$ , $\{1,2,4\}$  or $\{2,4\}$) is 
 \begin{eqnarray}
 \label{player_4_p3}
 \frac{1}{2}\Big(\frac{\lambda}{\lambda+3\lambda_0}\Big)^2 =   \frac{1}{2}\Big(\frac{1}{1+3\eta}\Big)^2.
 \end{eqnarray}

 Utility of player 2  if  it
 chooses   $\{2\}$ (or equivalently $\{1,2\}$) is 
 \begin{eqnarray}
  \label{player_4_p4}
  \Big(\frac{\lambda}{\lambda+3\lambda_0}\Big)^2 =   \Big(\frac{1}{1+3\eta}\Big)^2.
 \end{eqnarray}
 From equations \eqref{player_4_p3} and \eqref{player_4_p4}, player 2 gets strictly better utilities when he chooses $\{2\}$. Thus, comparing equation \eqref{player_4_p2} and \eqref{player_4_p4}, $\{1,2,3,4\}$  (or equivalently $\{2,3,4\}$) lies in best response of player 2, if
 \begin{eqnarray*}
 \frac{1}{3}\Big(\frac{1}{1+2\eta}\Big)^2 & \geq & \Big(\frac{1}{1+3\eta}\Big)^2 \mbox{ , i.e.,  if, } \\
 (1+3\eta)  & \geq & \sqrt{3}(1+2\eta) \mbox{ , i.e.,  if, } \\
 (3-2\sqrt{3})\eta & \geq & (\sqrt{3}-1).
 \end{eqnarray*}
which is not possible. Hence, a strategy profile leading to the $\mathcal{P}_2$ type partition (with three players in one coalition and one in other) cannot be a NE.

Now we consider BR of player 3 against $\{1\}$, $\{2\}$ and $\{1,2,3,4\}$ strategy of player 1,2 and 4 respectively.

Utility of player 3 when it chooses $\{1,2,3,4\}$ (or equivalently $\{2,3,4\}$, $\{3,4\}$ or $\{1,3,4\}$), is
\begin{eqnarray}
\label{player_4_partition2}
\frac{1}{2}\Big(\frac{\lambda}{\lambda+3\lambda_0}\Big)^2 =   \frac{1}{2}\Big(\frac{1}{1+3\eta}\Big)^2.
\end{eqnarray}
Utility of player 3 when it chooses $\{3\}$ (or equivalently $\{2,3\}$, $\{1,3\}$), is
\begin{eqnarray}
\label{player_4_partition3}
\Big(\frac{\lambda}{\lambda+4\lambda_0}\Big)^2 =   \Big(\frac{1}{1+4\eta}\Big)^2.
\end{eqnarray}

 Comparing equation \eqref{player_4_partition2} and \eqref{player_4_partition3}, $\{1,2,3,4\}$ (or equivalently $\{2,3,4\}$, $\{3,4\}$ or $\{1,3,4\}$) lies in best response of player 3, if
 \begin{eqnarray*}
  \frac{1}{2}\Big(\frac{1}{1+3\eta}\Big)^2 & \geq & \Big(\frac{1}{1+4\eta}\Big)^2 \mbox{ , i.e.,  if, } \\ 
  (1+4\eta)  & \geq & \sqrt{2}(1+3\eta) \mbox{ , i.e.,  if, } \\
  (4-3\sqrt{2})\eta  & \geq &  (\sqrt{2}-1).
 \end{eqnarray*}
which is not possible. Hence, a strategy profile leading to the $\mathcal{P}_3$ type partition cannot be a NE (as explained above).

Now the strategy profiles leading to TTC partition is remaining. Consider best response of player 1 against $\{1,2\},$\{3,4\} and $\{3,4\}$ strategy of player 2,3 and 4 respectively. 

Utility of player 1 if it chooses $\{1,2\}$ (or equivalently $\{1,2,3,4\}$, $\{1,2,3\}$ or $\{1,2,4\}$) is
\begin{eqnarray}
\frac{1}{2}\Big(\frac{\lambda}{\lambda+2\lambda_0}\Big)^2 =   \frac{1}{2}\Big(\frac{1}{1+2\eta}\Big)^2.
\label{util_ttc_4}
\end{eqnarray}
Utility of player 1 if it chooses $\{1\}$ (or equivalently $\{1,3,4\}$, $\{1,3\}$ or $\{1,4\}$) is
\begin{eqnarray}
\Big(\frac{\lambda}{\lambda+3\lambda_0}\Big)^2 =   \Big(\frac{1}{1+3\eta}\Big)^2.
\label{util_ttc1_4}
\end{eqnarray}
 Comparing equation \eqref{util_ttc_4} and \eqref{util_ttc1_4}, $\{1\}$ (or equivalently $\{1,3,4\}$, $\{1,3\}$ or $\{1,4\}$) lies in best response of player 1, if
 \begin{eqnarray*}
  \Big(\frac{1}{1+3\eta}\Big)^2 &\geq & \frac{1}{2}\Big(\frac{1}{1+2\eta}\Big)^2 \mbox{ , i.e.,  if, } \\
  \sqrt{2}(1+2\eta)  &\geq & (1+3\eta) \mbox{ , i.e.,  if, } \\
  (\sqrt{2}-1) &\geq & (3-2\sqrt{2})\eta \mbox{ , i.e.,  if, } \\
  2.414 &\geq & \eta.
 \end{eqnarray*}
Thus, TTC is a NE-partition when $\eta > 2.414$. Else, ALC is the NE-partition in this range.

\textbf{Case 2: When $0.67 <\eta \leq 0.75$}  In this range adversary becomes insignificant in ALC. Hence, only the last step in previous case is changed.

Utility of player 3 when it chooses $\{3\}$ (or equivalently $\{2,3\}$, $\{1,3\}$), is
\begin{eqnarray}
\label{player_4_partition6}
\frac{1}{4^2} &= & \frac{1}{16}.
\end{eqnarray}
 Comparing equation \eqref{player_4_partition2} and \eqref{player_4_partition6}, $\{1,2,3,4\}$ (or equivalently $\{2,3,4\}$, $\{3,4\}$ or $\{1,3,4\}$) lies in best response of player 3, if
 \begin{eqnarray*}
  \frac{1}{2}\Big(\frac{1}{1+3\eta}\Big)^2 & \geq &  \frac{1}{16} \mbox{ , i.e.,  if, } \\ 
 4  & \geq & \sqrt{2}(1+3\eta)\mbox{ , i.e.,  if, }  \\
  (4-\sqrt{2})  & \geq &  3\sqrt{2}\eta \mbox{ , i.e.,  if, } \\
  0.61 & \geq &  \eta.
 \end{eqnarray*}
 which is not possible. Hence, ALC$^o$ is the NE-partition in this range. TTC cannot be NE-partition in this range as one of the player can deviate and stay alone (since $\mathcal{P}_3$ type partitions are with significant adversary are possible in this range).

\textbf{Case 3: When $0.5 <\eta \leq 0.67$}
In this range, adversary becomes insignificant in ALC and any $\mathcal{P}_3$ type partition. GC is still not a NE-partition as explained in case 1 and hence player 1'st best response is $\{1\}$.

Consider best response of player 2 against $\{1\}$, $\{1,2,3,4\}$ and $\{1,2,3,4\}$ strategy of player 1,3 and 4 respectively.

Utility of player 2 when it chooses $\{1,2,3,4\}$ (or equivalently $\{2,3,4\}$), is given by equation \eqref{player_4_p2}.

From equation \eqref{min_util_mult_partition}, utility of player 2  if  it
 chooses   $\{2,3\}$ (or equivalently $\{1,2,3\}$ , $\{1,2,4\}$  or $\{2,4\}$) is 
 \begin{eqnarray}
 \label{player_4_pla3}
 \frac{1}{2}\frac{1}{3^2} &= &  \frac{1}{18}.
 \end{eqnarray}

 Utility of player 2  if  it
 chooses   $\{2\}$ (or equivalently $\{1,2\}$) is 
 \begin{eqnarray}
  \label{player_4_pla4}
\frac{1}{3^2}&=&    \frac{1}{9}.
 \end{eqnarray}
 From equations \eqref{player_4_pla3} and \eqref{player_4_pla4}, player 2 gets strictly better utilities when he chooses $\{2\}$. Thus, comparing equation \eqref{player_4_p2} and \eqref{player_4_pla4}, $\{1,2,3,4\}$  (or equivalently $\{2,3,4\}$) lies in best response of player 2, if
 \begin{eqnarray*}
 \frac{1}{3}\Big(\frac{1}{1+2\eta}\Big)^2 & \geq & \frac{1}{9} \\
 3 & \geq & \sqrt{3}(1+2\eta) \mbox{ , i.e.,  if, } \\
 (3-\sqrt{3}) & \geq & 2\sqrt{3}\eta \mbox{ , i.e.,  if, } \\
 0.366 & \geq & \eta.
 \end{eqnarray*}
which is not possible. Hence, a strategy profile leading to the $\mathcal{P}_2$ type partition (with three players in one coalition and one in other) cannot be a NE.

Now the strategy profiles leading to TTC partition is remaining. Consider best response of player 1 against $\{1,2\},$\{3,4\} and $\{3,4\}$ strategy of player 2,3 and 4 respectively. 

Utility of player 1 if it chooses $\{1,2\}$ (or equivalently $\{1,2,3,4\}$, $\{1,2,3\}$ or $\{1,2,4\}$) is given by equation \eqref{util_ttc_4}.

Utility of player 1 if it chooses $\{1\}$ (or equivalently $\{1,3,4\}$, $\{1,3\}$ or $\{1,4\}$) is given by \eqref{player_4_pla4}.

 Comparing equation \eqref{util_ttc_4} and \eqref{player_4_pla4}, $\{1\}$ (or equivalently $\{1,3,4\}$, $\{1,3\}$ or $\{1,4\}$) lies in best response of player 1, if
 \begin{eqnarray*}
 \frac{1}{9} &\geq & \frac{1}{2}\Big(\frac{1}{1+2\eta}\Big)^2 \mbox{ , i.e.,  if, } \\
  \sqrt{2}(1+2\eta)  &\geq & 3 \mbox{ , i.e.,  if, }  \\
  2\sqrt{2}\eta &\geq & (3-\sqrt{2}) \mbox{ , i.e.,  if, }  \\
  \eta &\geq &  0.567.
 \end{eqnarray*}
Thus, TTC and ALC$^o$ are the NE-partition in this range when $0.5 \leq \eta \leq 0.567$.

\textbf{Case 4: When $0 \leq \eta \leq 0.5$}
In this range, adversary is significant only in GC. 
Consider BR of player 1 against $\{1,2,3,4\}$, $\{1,2,3,4\}$ and $\{1,2,3,4\}$ strategy of player 2, 3 and 4 respectively. 
Utility of player 1 if it chooses $GC = \{1,2, 3,4\}$ (partition GC is formed) 
is given by \eqref{util_gc_4_adv}.
From equation \eqref{min_util_mult_partition}, utility of player 1  if  it
 chooses   $\{1,2,3\}$ (or equivalently $\{1,3,4\}$ or $\{1,2,4\}$) is 
\begin{eqnarray}
\label{utility_par3_4_adv}
\frac{1}{3} \frac{1}{2^2} & = & \frac{1}{12}.
\end{eqnarray}
Similarly,  utility of player 1  if  it
 chooses $\{1,2\}$  (or equivalently $\{1,3\}$ or $\{1,4\}$) is  
\begin{eqnarray}
\label{adv_util_plo1}
\frac{1}{2} \frac{1}{2^2} & = & \frac{1}{8}.
\end{eqnarray}
Also, utility of player 1  if  it
 chooses $\{1\}$ is  
\begin{eqnarray}
\label{adv_util_plo2}
\frac{1}{2^2} & = & \frac{1}{4}.
\end{eqnarray}
  From equations \eqref{utility_par3_4_adv}, \eqref{adv_util_plo1} and \eqref{adv_util_plo2}, player 1 gets strictly better utilities when he chooses $\{1\}$. Thus, comparing equation \eqref{util_gc_4_adv} and \eqref{adv_util_plo2}, $\{1,2,3,4\}$  lies in best response of player 1, if
\begin{eqnarray*}
 \frac{1}{4}\Big(\frac{1}{1+\eta}\Big)^2 & \geq &  \frac{1}{4} \mbox{ , i.e.,  if, } \\
 1  & \geq &  (1+\eta).
\end{eqnarray*}
which is not possible.

Next we consider BR of player 2 against $\{1\}$, $\{1,2,3,4\}$ and $\{1,2,3,4\}$ strategy of player 1,3 and 4 respectively.

Utility of player 2 when it chooses $\{1,2,3,4\}$ (or equivalently $\{2,3,4\}$), is
given by \eqref{utility_par3_4_adv}.

From equation \eqref{min_util_mult_partition}, utility of player 2  if  it
 chooses   $\{2,3\}$ (or equivalently $\{1,2,3\}$ , $\{1,2,4\}$  or $\{2,4\}$) is 
 \begin{eqnarray}
 \label{util_mult_adv_pol1}
 \frac{1}{2}\frac{1}{3^2} & = & \frac{1}{18}.
 \end{eqnarray}

 Utility of player 2  if  it
 chooses   $\{2\}$ (or equivalently $\{1,2\}$) is 
 \begin{eqnarray}
  \label{util_mult_adv_pol2}
 \frac{1}{3^2} & = & \frac{1}{9}.
 \end{eqnarray}
 From equations \eqref{utility_par3_4_adv}, \eqref{util_mult_adv_pol1} and \eqref{util_mult_adv_pol2}, player 2 gets strictly better utilities when he chooses $\{2\}$. 

Similarly, player 3 gets better utilities when he chooses to remain alone. Hence, we have ALC$^o$ as the NE-partition.

Now the strategy profiles leading to TTC partition is remaining. Consider best response of player 1 against $\{1,2\},$\{3,4\} and $\{3,4\}$ strategy of player 2,3 and 4 respectively.

Utility of player 1 if it chooses $\{1,2\}$ (or equivalently $\{1,2,3,4\}$, $\{1,2,3\}$ or $\{1,2,4\}$) is
\begin{eqnarray}
\label{ttc_util_pol1}
\frac{1}{2}\frac{1}{2^2} & =  \frac{1}{8}.
\end{eqnarray}

Utility of player 1 if it chooses $\{1\}$ (or equivalently $\{1,3,4\}$, $\{1,3\}$ or $\{1,4\}$) is
given by \eqref{util_mult_adv_pol2}. 

Comparing \eqref{ttc_util_pol1} and  \eqref{util_mult_adv_pol2} player 1's best response is $\{1,2\}$. Hence, TTC$^o$ is also a NE-partition.

\section*{Computations for $n\le4$ without adversary}
\label{comp_small_without_adv}

\subsection*{ When $n=2$:}
To compute NE, we consider \emph{best response} (BR) of player 1 against strategy $\{1,2\}$ of player 2.

From equation \eqref{Eqn_USm_player}, utility of player 1 when it chooses $\{1,2\}$, we have
\begin{eqnarray}
\label{util_gc_adv_2}
 \frac{1}{2}\frac{1}{1^2} =   \frac{1}{2}.
\end{eqnarray}
Similarly, utility of player 1 when it chooses $\{1\}$, we have
\begin{eqnarray}
\label{util_alc_adv_2}
\frac{1}{2^2} = \frac{1}{4}.
\end{eqnarray}
Comparing equation \eqref{util_gc_adv_2} and \eqref{util_alc_adv_2}, GC$^o$ is formed ,i.e., $\{1,2\}$ lies in best response of player 1 against $\{1,2\}$ strategy of player 2.Recall ALC is always a NE.
 The SO-partitions can be calculated from Lemma \ref{Lemma_SO_no_adv}.

\subsection*{ When $n=3$:}
To compute NE, we consider \emph{best response} (BR) of player 1 against strategy $\{1,2,3\}$ of player 2 and 3.

From equation \eqref{Eqn_USm_player}, utility of player 1 when it chooses $\{1,2,3\}$, we have
\begin{eqnarray}
\label{util_gc_noadv_2}
 \frac{1}{3}\frac{1}{1^2} =   \frac{1}{3}.
\end{eqnarray}
From equation \eqref{min_util_mult_partition} , utility of player 1 when it chooses $\{1,2\}$ (or equivalently $\{1,3\}$, we have
\begin{eqnarray}
\label{util_alc_noadve_2}
\frac{1}{2}\frac{1}{2^2} = \frac{1}{8}.
\end{eqnarray}
Similarly, utility of player 1 when it chooses $\{1\}$, we have
\begin{eqnarray}
\label{util_alc_noadv_2}
\frac{1}{2^2} = \frac{1}{4}.
\end{eqnarray}
Comparing equation \eqref{util_gc_noadv_2}, \eqref{util_alc_noadve_2} and \eqref{util_alc_noadv_2}, GC$^o$ is formed ,i.e., $\{1,2,3\}$ lies in best response of player 1 against $\{1,2,3\}$ strategy of player 2 and 3. Recall ALC$^o$ is always a NE. Also, $\mathcal{P}_2^o$
 type partitions can also be NE-partition for some strategy profiles. For example, $x_1 = \{1\}$, $x_2= \{2,3\}$ and $x_3= \{2,3\}$. The SO-partitions can be calculated from Lemma \ref{Lemma_SO_no_adv}.

\subsection*{ When $n=4$:}
To compute NE, we consider \emph{best response} (BR) of player 1 against strategy $\{1,2,3,4\}$ of player 2,3 and 4.

From equation \eqref{Eqn_USm_player}, utility of player 1 when it chooses $\{1,2,3,4\}$, we have
\begin{eqnarray}
\label{util_gc_noadv_4}
 \frac{1}{4}\frac{1}{1^2} =   \frac{1}{4}.
\end{eqnarray}
From equation \eqref{min_util_mult_partition} , utility of player 1 when it chooses $\{1,2,3\}$ (or equivalently $\{1,3,4\} or \{1,2,4\}$, we have
\begin{eqnarray}
\label{util_p2_noadv_4}
\frac{1}{3}\frac{1}{2^2} = \frac{1}{12}.
\end{eqnarray}
Similarly, utility of player 1 when it chooses $\{1,2\}$, we have
\begin{eqnarray}
\label{util_p3_noadv_4}
\frac{1}{2}\frac{1}{2^2} = \frac{1}{8}.
\end{eqnarray}
 utility of player 1 when it chooses $\{1\}$, we have
\begin{eqnarray}
\label{util_p4_noadv_4}
\frac{1}{2^2} = \frac{1}{4}.
\end{eqnarray}
Comparing equation \eqref{util_gc_noadv_4}, \eqref{util_p2_noadv_4},
\eqref{util_p3_noadv_4} and
\eqref{util_p4_noadv_4}, GC$^o$ is formed ,i.e., 
$\{1,2,3,4\}$ lies in best response of
 player 1 against $\{1,2,3,4\}$ strategy of player 2,3 and 4. 
Recall ALC$^o$ is always a NE. 
Also, TTC$^o$ type partitions can also be NE-partition for some strategy profiles. For example, $x_1 = \{1,2\}$, $x_2= \{1,2\}$, $x_3= \{3,4\}$ and $x_4 = \{3,4\}$. The SO-partitions can be calculated from Lemma \ref{Lemma_SO_no_adv}.}{}

%\newpage
%\bibliographystyle{abbrv}
%\bibliography{ref}
\end{document}

%% file: Intro1.tex
\begin{abstract}
 Cooperative game theory deals with systems where players want to cooperate to improve their payoffs. But players may choose coalitions in a non-cooperative manner, leading to a coalition-formation game. We consider such a game with several players (willing to cooperate) and an adamant player (unwilling to cooperate) involved in resource-sharing. Here, the strategy of a player is the set of players with whom it wants to form a coalition. Given a strategy profile, an appropriate partition of coalitions is formed;  players in each coalition maximize their collective utilities leading to a non-cooperative resource-sharing game among the coalitions, the utilities at the resulting equilibrium are shared via Shapley-value; these shares define the utilities of players for the given strategy profile in coalition-formation game. We also consider the utilitarian solution to derive the price of anarchy (PoA). 
 We considered a case with symmetric players and an adamant player; wherein we observed that players prefer to stay alone at Nash equilibrium when the number of players ($n$) is more than 4. In contrast, in majority of the cases, the utilitarian partition is grand coalition. Interestingly the PoA is smaller with an adamant player of intermediate strength. Further, PoA grows like $O(n)$.

% {\color{blue}The results are for identical players, but the coalition formation game framework is developed for  more   general cases with non-identical players.} 

%At the Nash equilibrium 
%    Taking these points into consideration, we observe that even when the total utility of players is higher in coalition but because of their selfish nature they end up with lower utility at equilibrium. In view of this, we also consider \emph{Price of Anarchy (PoA)} to estimate the loss of players because of their selfish behaviour.

\end{abstract}
\section{Introduction}
\label{Intro}
Resource sharing problem is a well-known problem that aims to find an optimal allocation of shared resources.  
We consider a scenario in which 
a common resource is to be shared amongst several users as in Kelly's mechanism \cite{kelly1997charging}; the utility  of any player is proportional to its bid and inversely proportional to the weighted sum   of  bids of all players, with the weights representing the influence factors.   This proportional allocation  problem is considered in a variety of other contexts;  e.g., \cite{stoica1996proportional} considers real-time performance in time-shared operating systems, \cite{kelly1998rate} considers rate allocation for communication networks, \cite{tun2019wireless} considers resource allocation  in  wireless  network  slicing, \cite{koutsopoulos2010auction} considers online auctions, etc.   We consider a similar game theoretic formulation with some important differentiating features:
i) possibility of cooperation among the willing players;  and,
ii) the presence of an adamant player, not interested in cooperation. Usually, the market giants (e.g., Amazon) %are not interested in forming coalitions and 
	tend to strive alone while smaller business entities (e.g., Flipkart, Walmart) try to look for  collaboration opportunities. %{\color{red}With players looking for such opportunities, we have a transferable utility cooperative game. } %The recent acquisition of Walmart by Flipart is one of the examples of resource sharing.}
 
%When players look out for cooperation opportunities, this leads to a cooperative game within the coalitions (e.g, \cite{narahari2014game}).

   Any   transferable utility cooperative game is defined by a set of players $N$ and the worth of each possible coalition  $\{\nu(S); S \subset  N\}$ (e.g, \cite{narahari2014game}, \cite{saad2009coalitional}). Majority of the analysis  related to cooperative games discuss the  emergence of  grand coalition (includes all players) as a successful partition and then consider the division of worth among the players;  Shapley value, Core  etc., are some  such solution concepts   (e.g, \cite{narahari2014game}).  But one can find many example scenarios, in which a partition of strict  coalitions (subsets) of $N$ might emerge out at some appropriate equilibrium (\cite{saad20091coalitional},\cite{saad2008distributed}).   In this context, one of the key challenges is to generate a partition, i.e., an exhaustive and disjoint division of the set of agents, such that the performance of the system is optimized (see for example, \cite{saad2009coalitional} and references therein). This leads to a utilitarian solution.  In contrast we consider a non-cooperative approach to generate partitions (e.g. as in \cite{Nevrekar2015ATO}, \cite{saad20091coalitional});  basically  the solution/partition would be stable against unilateral deviation. These are in general called as coalition formation games (CFGs) (\cite{saad2009coalitional}).

Another important aspect  of cooperative games is \textit{characteristic form games} and \textit{partition form games} \cite{saad2009coalitional}. Majority of the literature focuses on the former type of games and  a very little attention has been given to a  more general class of {\it partition form games}. In the former type of cooperative games, the  worth of a coalition depends only on the members of the coalition, 
 while in the latter type, the worth  is also influenced  by the partition of the  players outside the given coalition (\cite{saad2009coalitional}).
  These inter-coalitional dependencies, \TR{called externalities from coalition formation, }{}play a crucial role in many real-world   applications where agents have either conflicting or overlapping goals (e.g., \cite{yi2003endogenous}, \cite{hafalir2007efficiency}). Our problem falls into the latter category.

 We consider a CFG  in the presence of  an adamant player (not willing to cooperate) and seek for  a non-cooperative solution.  
In  our    game,  the     strategy of a  player   is the set of players with whom  it wants to form coalition as in \cite{Nevrekar2015ATO}.
Given the strategies of all players,  basically the preferences of all the players,  an  appropriate partition of   \emph{coalitions}  is  formed;  and players in each coalition maximize their collective utilities. This leads to a non-cooperative resource sharing game (RSG) among the coalitions.  The utilities at the resulting equilibrium are shared via Shapley value (confined to each coalition);   these shares define  the utilities of individual players for the given coalition suggestive preferences of all players in CFG. % for the given strategy profile.

We derived the solution of this non-cooperative CFG for the special case of  symmetric players (players with same influence factor).  %We demonstrated that players prefer to stay alone at Nash equilibrium (NE), when the  number of players ($n$) is more than 4. 
For smaller number of players ($n \leq 4$), the  partitions at NE  depend upon the relative strength of the adamant player and that of the others (referred as $\eta$). 
The partitions at NE are not monotone with $\eta$:  coarser partitions result at lower and higher values of  $\eta$, while we have finer partitions for  intermediate values.  This non-monotone behaviour is absent  for $n > 4$;  the players prefer to  remain alone at equilibrium (irrespective of $\eta$), i.e., the finermost partition emerges out.  
 
 We also consider the  utilitarian solution (maximizes the sum of utilities of all players) to derive the price of anarchy ($P_{oA}$), which captures the loss of players resulting due  to their rational behaviour.  The utilitarian solutions are also  non-monotone with $\eta$, however majority of times  grand coalition (coarsest partition) is the solution. For few cases a partition with two coalitions is the  solution.   
 Interestingly, the $P_{oA}$ is higher in the absence of  an adamant player even when the players do not share resources with  an extra player.  The $P_{oA}$ increases with   $n$ and  $\eta$.  It also increases  with decrease in  $\eta$  to zero. Interestingly, the limit in all the cases equals that in the system without  adamant player.

%% file: Social_optima_n_small.tex
{\small
\begin{table}[h]
\centering
\begin{tabular}{|c|c|c|c|c|}
\hline
  & Range                  & $\mathcal{P}$ at NE & $\mathcal{P}$ at SO & $P_{oA}$                                                                                                                  \\ \hline
1 & $\eta > \frac{n-1}{n}$ & ALC                 & GC                  & $\frac{1}{n}\Big(\frac{1+n\eta}{1+\eta}\Big)^2$ \\ \hline
2 &  $0.707 \le  \eta \leq \frac{n-1}{n}$                  &       ALC$^{o}$              &                    GC   &              $ \frac{n}{(1+\eta)^2}$                                                                                                       \\ \hline
3 &       $0.5 < \eta  \leq  0.707$                 &           ALC$^{o} $           &     $\mathcal{P}_2$              &                                                                                                                      $ \frac{2n}{(1+2\eta)^2}$ \\ \hline
4 &       $0.414 \le  \eta \leq 0.5$                 &          ALC$^{o} $           &      $\mathcal{P}_2^o$                & $\frac{n}{2}$                                                                                                                      \\ 
\hline 
%& &&& \\
5 &       $0 < \eta  \leq  0.414$                 &        ALC$^{o} $             &  GC                  &      \hspace{-3mm}$ \frac{{n}^{\hspace{1mm}}}{(1+\eta)^2}$                                                                                                                                                                                                                   \\ \hline
\end{tabular}
\vspace{1mm}
\caption{NE-partitions, SO-Partitions and $P_{oA}$ for $n > 4$}
\label{tab_PoA_large}
\vspace{-6mm}
\end{table}}
%\vspace{-1mm}
\section{Small number of players, $n \le 4$}
\label{sec:small_no}
%\vspace{-1mm}
In this section, we identify the  NE-partitions and derive the $P_{oA}$,    for $n \leq 4$, by direct computations. 
%{\color{red}GC insignificance}
\subsubsection*{When $n = 2$}
Here,  GC and ALC (or ALC$^o$) are the only possible partitions. %The ALC partition is equivalent to $\mathcal{P}_2$.
Some strategy profiles and the corresponding partitions can be seen from  Table \ref{tab:n=2}.
\vspace{-2mm}
\begin{table}[h]
\centering
\begin{minipage}{4cm}
	\hspace{-5mm}
	\centering
\begin{tabular}{|c|c|c|c|}
\hline
 $x_1$   & $x_2$   & $\mathcal{P}$         \\ \hline
GC& GC & GC     \\ \hline
ALC   & GC &ALC \\ \hline
\TR{ALC   & ALC   & ALC \\ \hline}{}
\end{tabular}
\vspace{1mm}
\caption{Partitions  at $n=2$}
\label{tab:n=2}
\end{minipage}
\begin{minipage}{4cm}
	\hspace{-5mm}
	\centering
\begin{tabular}{|c|c|c|c|c|}
	\hline
	$x_1$     & $x_2$     & $x_3$     & $\mathcal{P}$                                       \\ \hline
	\TR{ GC & GC & GC & GC                                \\ \hline }{}
	\{1,2\}   &GC&GC & \{\{1,2\},\{3\}\} \\
	&                 &                 & \{\{1\},\{2,3\}\}  \\ \hline
	ALC     &GC& GC &\{\{1\},\{2,3\}\}                            \\ \hline
	\TR{ ALC    &ALC   &GC & ALC                        \\ \hline}{}
	\TR{ALC    & ALC   & ALC     & ALC                         \\ \hline}{}
\end{tabular}
\vspace{0.5mm}
\caption{Partitions at $n=3$}
\label{tab:n=3}
\end{minipage}
\vspace{-6mm}
\end{table}

We begin with  deriving  the best responses (BR). Consider the case with    $\eta \ge 0.707$. Then from \eqref{Eqn_USm_player}, BR of player 2 against    player 1's  strategy, $x_1 = \{1,2\}$ is GC, because:
\vspace{-1mm}
$$
\frac{1}{2}\Big(\frac{\lambda}{\lambda+\lambda_0}\Big)^2   \geq   \Big(\frac{\lambda}{\lambda+ 2 \lambda_0}\Big)^2.
$$    Thus  both GC and ALC  are  NE-partitions  when  $\eta \ge 0.707$. In a similar way one can verify that the only NE-partition is ALC for 
 $0.5 < \eta \le  0.707 $ (see Table \ref{tab:my-table2}).
%\vspace{-1mm}
 \begin{table}[h]
\centering
\begin{tabular}{|c|c|c|c|c|}
\hline
 &  $\mathcal{P}$ at NE                   &  Range of parameters &  $\mathcal{P} \text{ at SO}$    &  $P_{oA}$                                                        \\ \hline
1     &  GC        & $\eta \ge 0.707$ & GC & $ \frac{1}{2} \Big( \frac{1+2\eta}{1+\eta} \Big)^2$ \\ 
       &  ALC      &        &         &                   \\  \hline
2     & ALC &$0.5 < \eta \leq 0.707$  &  $\mathcal{P}_2 $ & 1 \\ \hline
3 & ALC$^o$  & $0.414 \leq \eta \leq 0.5$  &   $\mathcal{P}_2 ^o$& 1 \\ \hline
4 & GC& $0 < \eta \leq 0.414$  & GC  & $\frac{2}{(1+\eta)^2}$ \\ 
      &  ALC$^o$      &                           &         &                   \\  \hline
\end{tabular}
\vspace{1mm}
\caption{NE-partitions, SO-partitions and $P_{oA}$ For  $n=2$ }
\label{tab:my-table2}
\vspace{-7mm}
\end{table}
 
 When  $0.414 \le  \eta \le  0.5$,   the adamant player is insignificant (gets 0 at NE) and ALC$^o$ is the  unique NE-partition. Interestingly, below $\eta \le 0.414$, 
the C-players find it beneficial  (again) to cooperate, note GC is also a NE. Thus we observe interesting  {\it non-monotone phenomenon  } with ratio of influence factors, $\eta$. 

\subsubsection*{When $n = 3$} 
In this case, we can have three types of partitions:   GC,  ALC  and ${\cal P}_2$ type partitions.  In any ${\cal P}_2$ type partition,  two of the C-players are together  in one coalition, while  the remaining  one is   alone.   
Some strategy profiles and their partitions are  in  Table \ref{tab:n=3}.
\Cmnt{\vspace{-2mm}
\begin{table}[h]
\centering
\begin{tabular}{|c|c|c|c|c|}
\hline
$x_1$     & $x_2$     & $x_3$     & $\mathcal{P}$                                       \\ \hline
\TR{ GC & GC & GC & GC                                \\ \hline }{}
\{1,2\}   &GC&GC & \{\{1,2\},\{3\}\} \\
               &                 &                 & \{\{1\},\{2,3\}\}  \\ \hline
 ALC     &GC& GC &\{\{1\},\{2,3\}\}                            \\ \hline
\TR{ ALC    &ALC   &GC & ALC                        \\ \hline}{}
\TR{ALC    & ALC   & ALC     & ALC                         \\ \hline}{}
\end{tabular}
\caption{Partitions for some strategy profiles for $n=3$}
\label{tab:n=3}
\vspace{-6mm}
\end{table} }

We derive the analysis by directly computing the BRs as in the previous case.  \TR{For example
consider the case with $\eta \ge 2.732$. It is easy to verify the following: when players  2 and 3 
choose $GC = \{1,2,3\}$, from \eqref{Eqn_USm_player} the utility of player 1  if  it
 chooses  $\{1,2, 3\}$ (partition GC is formed) 
is 
$$
\frac{1}{3}\Big(\frac{\lambda}{\lambda+\lambda_0}\Big)^2   >    \Big(\frac{\lambda}{\lambda+ 2 \lambda_0}\Big)^2,
$$ the utility of  the same player if   it chooses to be alone.  Thus, GC is a NE-partition when  $\eta \geq 2.732$. Using similar computations we observe that any  ${\cal P}_2$ and ALC  are  also NE-partitions in this range.  However one can easily compute  and observe that the utilities of C-players  at the first NE (GC) is bigger than that in the remaining two, thus GC is a preferred NE in this range. It is also the SO-partition.  However   $P_{oA}$ compares the SO  utilities  with those at worst NE and hence we have   $P_{oA}$ strictly bigger than one (first row in Table \ref{tab:my-table1}). 

\Cmnt{

Next, consider the case with $ 2.414    < \eta \le  2.732  $, and it is easy to verify the following: when player 1 and 3 chooses $\{1\}$ and $\{1,2,3\}$ respectively, from \eqref{Eqn_USm_player} the utility of player 2  if  it chooses  $\{1,2,3\}$ (TOC is formed) is 
$$
\frac{1}{2}\Big(\frac{\lambda}{\lambda+2\lambda_0}\Big)^2   >    \Big(\frac{\lambda}{\lambda+ 3 \lambda_0}\Big)^2,
$$ the utility of  the same player if   it chooses to be alone. Thus, both TOC and ALC are NE-partitions when  $ 2.414    < \eta \le  2.732  $. In a similar way one can verify that the only NE partition is ALC for  $0.67 <\eta \le 2.414 $.

In the range $0.5 < \eta \leq 0.67 $ and $\eta < 0.5$ adversary gets zero utility when partition with 4 and  3 coalitions is formed respectively. 
}
One can compute the required details for the remaining cases and the} {The} results
are summarized in Table \ref{tab:my-table1}.
\vspace{-1mm}
\begin{table}[h]
\centering
\begin{tabular}{|c|c|c|c|c|}
\hline
 &  $\mathcal{P}$ at NE                   &  Range&  $\mathcal{P}$ \text{at SO}    &$P_{oA}$                    \\ \hline
1     & GC        &   &  & \\ 
       & $\mathcal{P}_2$             &           $\eta \geq 2.732$      &  GC            &    $\frac{1}{3} \Big( \frac{1+3\eta}{1+\eta} \Big)^2$       \\       
              & ALC             &                           &          &  \\       \hline
2     &  $\mathcal{P}_2$  & $ 2.414   \leq \eta \le  2.732 $                                        & GC                          & $\frac{1}{3} \Big( \frac{1+3\eta}{1+\eta} \Big)^2$ \\ 
        &  ALC &  &  &  \\   \hline
3     &   ALC &   $ 0.707  \le   \eta  \le2.414$    &  GC & $\frac{1}{3} \Big( \frac{1+3\eta}{1+\eta} \Big)^2$   \\ \hline
4     &   ALC &   $   0.57\leq\eta \leq 0.707 $    &   $\mathcal{P}_2$ & $ \frac{2}{3} \Big( \frac{1+3\eta}{1+2\eta} \Big)^2$ \\ \hline
5     & $\mathcal{P}_2$  & $0.5 < \eta \leq 0.57 $  &  $\mathcal{P}_2$   & $ \frac{6}{(1+2\eta)^2}$  \\ 
 & ALC$^o$             &                           &          &  \\       \hline
6     & $\mathcal{P}_2^o$  & $0.414 \le \eta \leq 0.5$  & $\mathcal{P}_2^o$   &  $\frac{3}{2}$ \\ 
& ALC$^o$             &                           &          &  \\       \hline
7     & $\mathcal{P}_2^o$  & $0.15 \leq \eta \leq 0.414$  & GC   &  $\frac{3}{(1+\eta)^2}$ \\ & ALC$^o$             &                           &          &  \\       \hline
8     & GC  &   &   &     \\ 
&$\mathcal{P}_2^o$            &      $0 < \eta \leq 0.15$                     &      GC      & $\frac{3}{(1+\eta)^2}$  \\      
& ALC$^o$             &                           &          &  \\       \hline
\Cmnt{5   &   ALC  & $0 \leq\eta \leq 0.408 $ & ALC \\ \hline}
\end{tabular}
\vspace{1mm}
\caption{NE-partitions, SO-partitions and $P_{oA}$ For  $n=3$ }
\label{tab:my-table1}
\vspace{-2mm}
\end{table}
Important observations are: a) If GC is a NE-partition, all others are also NE-partitions; b) recall  ALC/ALC$^o$ is always a NE-partition; 
c)  the utilities  of all the players  at GC are bigger than  those at ${\cal P}_2$ or ALC, when GC is a NE-partition, thus GC is the preferred NE  in row 1 and 8 of Table \ref{tab:my-table1}; d) the  
 utilities  of all  players  at ${\cal P}_2$   are  bigger than  those at ALC,   when  ${\cal P}_2$   is an NE, in such cases,    ${\cal P}_2$ is the preferred one, etc.

\subsubsection*{When $n = 4$}
In this case, we can have four types of partitions: GC, ALC, $\mathcal{P}_2$ and $\mathcal{P}_3$ type partitions.  In any ${\cal P}_3$ type partition,  two of the C-players are together  in one coalition, while  the remaining two players  are alone. While ${\cal P}_2$ type partition can have either two players in each coalition or three players in one coalition and the remaining one is alone. We refer the first one  as TTC (partition with Two-Two  coalitions). Some of the strategy profiles and the corresponding partitions can be seen from Table \ref{tab:n=4}.

{\tiny
\begin{table}[h]
\vspace{-2mm}
\centering
\small
\begin{tabular}{|c|c|c|c|c|c|}
\hline
$x_1$       & $x_2$       & $x_3$       & $x_4$       & $\mathcal{P}$                                                   \\ \hline
\TR{
GC & GC & GC &GC & GC                                            \\   \hline}{}

 \{1,2,3\}     & GC & GC & GC & \{\{1,2,3\},\{4\}\}  \\
          &             &            &             & \{\{1\},\{2,3,4\}\}         \\ \hline
 \{1,2\}     & GC & GC & GC & \{\{1,2\},\{3,4\}\}  \\
           &             &            &             & \{\{1\},\{2,3,4\}\}         \\ \hline
\TR{ALC    & GC, &GC & GC& \{\{1\},\{2,3,4\}\}                    \\ \hline
ALC   & \{2,3\}     & GC & GC & \{\{1\},\{2,3\},\{4\}\} \\
            &             &             &            & \{\{1\},\{2\},\{3,4\}\} \\ \hline
ALC      & ALC      &GC & GC  & \{\{1\},\{2\},\{3,4\}\}                                   \\      \hline

 ALC     & ALC     & ALC      & GC& ALC            \\ \hline}{}
\end{tabular}
\vspace{1mm}
\caption{partitions at $n=4$}
\label{tab:n=4}
\vspace{-7mm}
\end{table}}

Once again BRs are computed directly  and the results are  in Table \ref{tab:my-table}.
\TR{ For example consider the case with    $\eta \geq 2.414 $. It is easy to verify the following:   when   players 2,3,4 choose $\{1,2\}$, $\{3,4\}$ and $\{3,4\}$ respectively, from \eqref{Eqn_USm_player} the utility of player 1  if  it chooses  $\{1,2\}$ (partition TTC is formed) is 
\begin{equation*}
\frac{1}{2}\Big(\frac{\lambda}{\lambda+2\lambda_0}\Big)^2   >    \Big(\frac{\lambda}{\lambda+ 3 \lambda_0}\Big)^2,
\end{equation*}
 the utility of  the same player if   it chooses to be alone. Thus TTC is a NE-partition  when  $\eta \geq 2.414$. Using similar computations we observe that ALC is also NE-partitions in this range. However one can easily compute  and observe that the utilities of C-players  at the first NE (TTC) is bigger than that in ALC, thus TTC is a preferred NE in this range. 
 From Lemma \ref{Lemma_SO_Partition},   GC is the SO-partition  in this range and   hence     $P_{oA}$ strictly bigger than one (first row in Table \ref{tab:my-table}). 
  
  In a similar way one can verify that the only NE-partition is ALC for 
 $0.5 < \eta  \le  2.414 $. One can also verify that the other partitions are not stable against unilateral deviation.
 
 In the range $0.57 \le \eta \leq 0.75 $, $0.5 < \eta \leq 0.57 $ and $\eta \le 0.5$ adamant player gets zero utility when partition with 5, 4 and  3 coalitions is formed respectively.
 
  This is summarized in Table \ref{tab:my-table}. 
 
 One can compute all the required details for the remaining cases and the results
are summarized in Table \ref{tab:my-table}.}{} Important observations are: a) GC is never a NE-partition; %b) recall  ALC is always a NE-partition; 
b)  the utilities  of all the players  at TTC are bigger than  those at ALC, when TTC is a NE-partition, thus TTC is the preferred NE.  
The non-monotone phenomena  observed for the case with $n=2$ can also be seen  for $n=3$ and $n=4$.

\begin{table}[h]
\vspace{-1mm}
\centering
\begin{tabular}{|c|c|c|c|c|}
\hline
& $\mathcal{P}$ at NE                 & Range   & $\mathcal{P}$ at SO & PoA\\ \hline
1     &  TTC                               &$\eta \geq 2.414$ & GC & $ \frac{1}{4} \Big( \frac{1+4\eta}{1+\eta} \Big)^2 $\\ 
  & ALC             &                           &          &  \\       \hline
2     &ALC                   & $0.75 \leq \eta\leq 2.414$   & GC&  $ \frac{1}{4} \Big( \frac{1+4\eta}{1+\eta} \Big)^2 $ \\ \hline
3     &     ALC$^o$      & $0.707 \leq \eta \le 0.75$  &GC &  $\frac{4}{(1+\eta)^2}$\\       \hline
4     &    ALC$^o$       & $0.57 \leq \eta \leq 0.707$  &$\mathcal{P} _2$  &$\frac{8}{(1+2\eta)^2}$  \\  \hline
5    &       TTC           & $0.5<  \eta \leq 0.57$   &  $\mathcal{P} _2$  & $\frac{8}{(1+2\eta)^2}$   \\ 
& ALC$^o$            &                           &          &  \\       \hline
6 &TTC$^o$ & $0.414 \le \eta \leq 0.5$  &  $\mathcal{P} _2^o$ & 2\\ 
& ALC$^o$            &                           &          &  \\       \hline
7 & TTC$^o$  & $0 < \eta \leq 0.414$  &  GC& $\frac{4}{(1+\eta)^2}$\\ 
&    ALC$^o$         &                           &          &  \\       \hline

\Cmnt{
3     & TTC, ALC                    & $0.408 \lambda \geq\lambda_0 \geq 0.354 \lambda$   &     $\mathcal{P}_3$\\  \hline
4     &  ALC          & $0.354 \lambda \ge\lambda_0 \ge 0.289\lambda$ &     $\mathcal{P}_3$\\ \hline
5     &  ALC          & $0.289\lambda \ge\lambda_0 \ge 0$ &  ALC \\ \hline}
\end{tabular}
\vspace{1mm}
\caption{NE-partitions, SO-partitions and $P_{oA}$ For  $n=4$}
\label{tab:my-table}
\vspace{-6mm}
\end{table}

%% file: Conclusions.tex
\section{Conclusions}
 
We consider a  coalition formation game with players exploring cooperation opportunities in a non-cooperative manner, where the utilities of players/coalitions are resultant of a resource sharing game.  We developed a framework 
to study the partitions  (non-overlapping and exhaustive set of  coalitions) that emerge at equilibrium. The  strategy of a player  is the set of players with whom it wants to form coalition, while the utilities of  players  are defined via (Shapley values of) the utilities of their coalitions and these coalitions/partition is formed based on the choice of all players;  the resulting coalitions     involve in a non-cooperative game along with an adamant player (not willing to cooperate) and the utilities at the equilibrium define the utilities of the coalitions.

 Our primary aim is to identify the NE-partitions, we also derive the partitions that result at utilitarian solution (maximizes the sum of utilities).   We observe that the agents 
derive much lower utilities at NE than that  at utilitarian solution,  and this loss is because of their strategic behaviour.  We considered $P_{oA}$ (price of anarchy) to estimate the loss. 

We considered a case study   with  symmetric players (with and without  adamant player),   and  have  the following important  results: 
i)  when $n > 4$,  the players prefer to stay alone  at NE, while their preferences are  coarser partitions (either grand coalition or a partition with two coalitions) at utilitarian solution;   ii) players derive  larger  utilities  at coarser (with smaller number of coalitions) partitions, if the latter emerges at equilibrium;   iii) thus the price of anarchy is significantly high, in fact increases as $O(n).$ 
iii) We see a non-monotone behaviour in the NE-partitions when $n \le 4$, with  the  strength of the adamant player measured via the ratio of influence factors; and   this behaviour is primarily because the adamant player gets insignificant at  more number of partitions, as its strength reduces;
%iii) the better partitions (defined as in \eqref{Eqn_Max_Coalition_rule_0}) provide better utilities for each of the players if they are NE-partitions and hence can be preferred;  
iv) the  $P_{oA}$  is smaller when the adamant player is of intermediate strength; it increases either as the strength of adamant player increases or as the strength reduces to zero; the limit in both the cases equals that in the system without  adamant player.